\RequirePackage{lineno}
\documentclass[twocolumn,aps,prc,showpacs,superscriptaddress,floatfix]{revtex4-2}
\usepackage{rotating}
\usepackage{epsfig} 
\usepackage{hyperref}
\usepackage{color}
\usepackage{lineno}
\usepackage{url}
\usepackage{multirow}

\usepackage{amsmath}
\usepackage{graphicx}
\usepackage[caption=false]{subfig}

\graphicspath{ {./plot/}, {./image/} }
\DeclareGraphicsExtensions{ .pdf, .png}
\makeatletter
\renewcommand{\p@subsection}{}
\renewcommand{\p@subsubsection}{}
\makeatother

\makeatletter
\let\LN@equation\equation
\let\LN@endequation\endequation
\renewcommand{\equation}{\linenomath\LN@equation}
\renewcommand{\endequation}{\LN@endequation\endlinenomath}
\let\LN@gather\gather
\let\LN@endgather\endgather
\renewcommand{\gather}{\linenomath\LN@gather}
\renewcommand{\endgather}{\LN@endgather\endlinenomath}
\makeatother

\begin{document}

%------------------------------------------------------------------------------------------------%

\newcommand{\der}{\text{d}}
\newcommand{\rp}{\Psi_{\textsc{rp}}}

\newcommand{\qcd}{{\textsc{qcd}}}
\newcommand{\pos}{{\textsc{os}}}
\newcommand{\pss}{{\textsc{ss}}}
\newcommand{\cme}{{\textsc{cme}}}
\newcommand{\ampt}{{\textsc{ampt}}}
\newcommand{\hijing}{{\textsc{hijing}}}
\newcommand{\poi}{{\textsc{poi}}}
\newcommand{\snn}{\sqrt{s_{\textsc{nn}}}}
\newcommand{\srp}{{\textsc{rp}}}
\newcommand{\ssp}{{\textsc{sp}}}
\newcommand{\spp}{{\textsc{pp}}}
\newcommand{\sep}{{\textsc{ep}}}
\newcommand{\dg}{\Delta\gamma}
\newcommand{\enf}{\epsilon_{\rm nf}}
\newcommand{\fcme}{f_{\textsc{cme}}}
\newcommand {\mean}[1]   {\langle{#1}\rangle}

\newcommand {\red}[1]   {\textcolor{red}{#1}}
\newcommand {\blue}[1]  {\textcolor{blue}{#1}}
\newcommand {\green}[1] {\textcolor{green}{#1}}

\newcommand{\fullnumber}{(4.1 \pm 1.4 \pm 4.6)\%}
\newcommand{\subnumber}{(-4.0 \pm 1.7 \pm 2.1)\%}
\newcommand{\sssnumber}{(-5.0 \pm 1.9 \pm 1.8)\%}
\newcommand{\fullnumberlowpt}{(6.8 \pm 3.0 \pm 5.5)\%}
\newcommand{\subnumberlowpt}{(-3.4 \pm 3.5 \pm 2.5)\%}
\newcommand{\sssnumberlowpt}{(-5.7 \pm 3.9 \pm 1.9)\%}
\newcommand{\fullnumberhighpt}{(-5.9 \pm 3.6 \pm 4.8)\%}
\newcommand{\subnumberhighpt}{(-8.7 \pm 4.5 \pm 3.2)\%}
\newcommand{\sssnumberhighpt}{(-10.0 \pm 5.1 \pm 3.0)\%}
\newcommand{\nosclfullnumber}{(6.5 \pm 1.7)\%}
\newcommand{\nosclsubnumber}{(-3.2 \pm 2.0)\%}
\newcommand{\nosclsssnumber}{(-4.4 \pm 2.2)\%}
\newcommand{\nosclfullnumberlowpt}{(9.9 \pm 3.6)\%}
\newcommand{\nosclsubnumberlowpt}{(-1.5 \pm 4.3)\%}
\newcommand{\nosclsssnumberlowpt}{(-4.6 \pm 4.7)\%}
\newcommand{\nosclfullnumberhighpt}{(-3.5 \pm 4.5)\%}
\newcommand{\nosclsubnumberhighpt}{(-6.3 \pm 5.6)\%}
\newcommand{\nosclsssnumberhighpt}{(-8.2 \pm 6.2)\%}
\newcommand{\simufullnumber}{(5.1 \pm 1.7)\%}
\newcommand{\simusubnumber}{(-4.3 \pm 2.0)\%}
\newcommand{\simusssnumber}{(-5.6 \pm 2.2)\%}
\newcommand{\simufullnumberlowpt}{(8.4 \pm 3.6)\%}
\newcommand{\simusubnumberlowpt}{(-2.7 \pm 4.2)\%}
\newcommand{\simusssnumberlowpt}{(-5.7 \pm 4.6)\%}
\newcommand{\simufullnumberhighpt}{(-5.5 \pm 4.4)\%}
\newcommand{\simusubnumberhighpt}{(-8.0 \pm 5.5)\%}
\newcommand{\simusssnumberhighpt}{(-9.9 \pm 6.2)\%}
\newcommand{\eampt}{(10 \pm 2 {\rm (stat.)})\%} 
\newcommand{\senf}{0.7} % \senf = 0.07 / \eampt
\newcommand{\fitdataeamptratio}{2.08\times\text{cent}+0.25} % x means centrality, 50%=0.5
\newcommand{\fitdataenfratio}{1.86\times\text{cent}+0.24}

%------------------------------------------------------------------------------------------------%

%\linespread{1.6}
\title{Two- and three-particle nonflow contributions to the chiral magnetic effect measurement by spectator and participant planes in relativistic heavy ion collisions}

\author{Yicheng Feng}
\email{feng216@purdue.edu}
\address{Department of Physics and Astronomy, Purdue University, West Lafayette, IN 47907, USA}

\author{Jie Zhao}
\email{zhao656@purdue.edu}
\address{Department of Physics and Astronomy, Purdue University, West Lafayette, IN 47907, USA}

\author{Hanlin Li}
\email{lihl@wust.edu.cn}
\address{College of Science, Wuhan University of Science and Technology, Wuhan, Hubei 430065, China}

\author{Hao-jie Xu}
\email{haojiexu@zjhu.edu.cn}
\address{School of Science, Huzhou University, Huzhou, Zhejiang 313000, China}

\author{Fuqiang Wang}
\email{fqwang@purdue.edu}
\address{Department of Physics and Astronomy, Purdue University, West Lafayette, IN 47907, USA}
\address{School of Science, Huzhou University, Huzhou, Zhejiang 313000, China}

\date{\today} %this is useful in drafting stage
%\draftversion{12}

%------------------------------------------------------------------------------------------------%

\begin{abstract}
    Correlation measurements with respect to the spectator and participant planes in relativistic heavy ion collisions were proposed to extract the chiral magnetic effect (\cme) from background dominated azimuthal correlators. This paper investigates the effects of two- and three-particle nonflow correlations on the extracted \cme\ signal fraction, $\fcme$. It is found, guided by a multiphase transport (\ampt) model and the heavy ion jet interaction generator (\hijing) together with experimental data, that the nonflow effects amount to approximately $(4\pm5)$\% and $(-5\pm3)$\% %$\fullnumber$ and $\sssnumber$ 
    without and with pseudorapidity gaps, respectively, in 20-50\% centrality Au+Au collisions at $\snn= 200 \text{ GeV}$.

\end{abstract}


%------------------------------------------------------------------------------------------------%

\maketitle

%------------------------------------------------------------------------------------------------%

\section{Introduction}

In quantum chromodynamics (\qcd) vacuum, topological charge fluctuations can cause chiral anomaly in local domains, which violates the $\mathcal{CP}$ symmetry~\cite{Lee:1974ma,Morley:1983wr,Kharzeev:1998kz,Kharzeev:2007jp,Fukushima:2008xe}.
Because the spin of quarks is either parallel or anti-parallel to strong magnetic field depending on their charge, such a chiral anomaly would result in charge separation along the magnetic field. 
This is called the chiral magnetic effect (\cme)~\cite{Kharzeev:1998kz,Kharzeev:2007jp,Fukushima:2008xe}.
In non-central heavy ion collisions, the overlap participant zone allows the formation of metastable topological domains, whereas the spectator protons can provide an intense, transient magnetic field perpendicular on average to the reaction plane (\srp, spanned by the impact parameter and beam directions)~\cite{Kharzeev:2007jp,Bzdak:2011yy,Huang:2015oca}.
Thus, the \cme\ is expected in relativistic heavy ion collisions, and,
if measured, would be a strong evidence for local $\mathcal{CP}$ violation
in \qcd~\cite{Kharzeev:2004ey}.

Charge-dependent azimuthal correlators~\cite{Voloshin:2004vk} are used to measure the \cme-induced charge separation, $\dg=\gamma_{\textsc{os}}-\gamma_{\textsc{ss}}$, where
\begin{equation} \label{gamma}
    \gamma_{\alpha\beta}=\mean{\cos(\phi_{\alpha}+ \phi_{\beta}-2\rp)}\,,
\end{equation}
$\phi$ is the azimuthal angle of particle of interest (\poi), the subscript $\alpha,\beta$ indicate charge signs (\textsc{os} for opposite-sign and \textsc{ss} for same-sign pairs) of two different particles, and $\rp$ is the \srp\ azimuthal angle.
Strong positive $\Delta\gamma$ signals have been observed in both large collision systems (Au+Au at RHIC~\cite{Abelev:2009ac,Abelev:2009ad,Adamczyk:2014mzf,Adamczyk:2013hsi} and 
Pb+Pb at the LHC~\cite{Abelev:2012pa,Khachatryan:2016got,Sirunyan:2017quh,Acharya:2017fau,Acharya:2020rlz}) and small systems (d+Au at RHIC~\cite{STAR:2019xzd} and %Zhao:2017wck,Zhao:2017ckp
p+Pb at the LHC~\cite{Khachatryan:2016got,Sirunyan:2017quh}).
No \cme\ signal is expected in the latter, indicating large background contaminations in $\Delta\gamma$.
The backgrounds are caused by two-particle (2p) nonflow correlations, such as resonance decays, coupled with elliptic flow ($v_2$) of the correlated pairs~\cite{Voloshin:2004vk,Wang:2009kd,Bzdak:2009fc,Schlichting:2010qia,Wang:2016iov,Zhao:2019hta}. It can be expressed as
\begin{equation}
    \Delta\gamma_{\rm bkgd} = \frac{N_{\rm 2p}}{N^2} \langle\cos(\phi_\alpha+\phi_\beta-2\phi_{\rm 2p})\rangle v_{2,{\rm 2p}}\,,
    \label{eq:bkgd}
\end{equation}
where $N$ is the \poi\ multiplicity of a single charge ($N\approx N_+\approx N_-$), $N_{\rm 2p}$ is the number of correlated pairs (such as the parent resonances), $\phi_{\rm 2p}$ is the azimuthal angle of the parent, and $v_{2,{\rm 2p}}$ is its elliptic flow w.r.t.~\srp, $v_{2,{\rm 2p}}=\mean{\cos2(\phi_{\rm 2p}-\rp)}$. 
Note Eq.~(\ref{eq:bkgd}) refers to the difference between \pos\ and \pss\ (i.e.~{\em charge-dependent}); the sign-independent effects are already canceled. 
This background is a combined effect of 2p nonflow correlations and flow, and is customarily called ``flow-induced" background.

To suppress the backgrounds, many techniques have been exploited, such as event shape engineering~\cite{Schukraft:2012ah,Adamczyk:2013kcb,Sirunyan:2017quh,Acharya:2017fau} and differential measurements in invariant mass~\cite{Zhao:2017nfq,Adam:2020zsu}. %Zhao:2017wck,
These studies indicate that the backgrounds are dominant and the possible \cme\ signal is consistent with zero.
Recently, a new method~\cite{Xu:2018prl,Xu:2017qfs} was invented to extract the \cme\ signal by comparing $\Delta\gamma$ and $v_{2}$ with respect to two {\em different} planes (instead of the unmeasurable \srp\ in Eq.~(\ref{gamma})) -- the participant plane (\spp, reconstructed from produced particles) and the spectator plane (\ssp, reconstructed from spectators).
We refer to the method as the \ssp/\spp\ method. 
The method is, arguably, the most robust one on market as it measures two quantities in the {\em same} event which contain different amounts of the \cme\ signal and background. It would give a rather robust measure of the \cme\ if there were no nonflow contaminations. Such measurements have been performed by the STAR Collaboration~\cite{Zhao:2018blc,Zhao:2020utk}. The purpose of this paper is to investigate the effects of nonflow contaminations on the extracted \cme\ signal.

The rest of the paper is organized as follows.
Section~\ref{sec:method} recaps the \ssp/\spp\ method and introduces a notation scheme to signal quantities that are affected by nonflow.
Section~\ref{sec:nonflow} describes the nonflow contaminations to $v_2^*$ and $\dg^*$ (or $C_3^*$).
Section~\ref{sec:model} presents model simulation results by A Multiphase Transport (\ampt) model and the Heavy Ion Jet Interaction Generator (\hijing), respectively.
Section~\ref{sec:realdata} makes quantitative estimations of nonflow contributions to the $\fcme^*$ in real data analysis by combining \ampt\ and \hijing\ simulation results together with inputs from experimental data.
Section~\ref{sec:summary} summarizes our work and gives a brief outlook.

%------------------------------------------------------------------------------------------------%

\section{Methodology}\label{sec:method}

The $\ssp/\spp$ method~\cite{Xu:2017qfs} is rather straightforward.
It exploits fluctuations in the collision geometry~\cite{Alver:2006wh,Alver:2008zza} that yields a non-unity $a\equiv\mean{\cos2(\Psi_\spp-\Psi_\ssp)}$.
Since the \cme\ backgrounds are induced by $v_2$ of particles in the participant zone, $\dg$ w.r.t.~\spp\ should contain the maximal background. As the background strength is proportional to $v_2$ (see Eq.~(\ref{eq:bkgd})), it is reduced by the factor of $a$ along the \ssp.
On the other hand, since the \cme\ signal is along the magnetic field created mainly by the spectator protons, $\dg$ w.r.t.~\ssp\ should contain the maximal signal. 
Its signal strength along the \spp\ would be reduced by, presumably, the same factor of $a$ based on the definition of the $\dg$ variable.
In other words, in the same event, the \cme\ signal is projected from \ssp\ to \spp, and the background is projected from \spp\ to \ssp.
With simple algebra, the \cme\ signal fraction in $\Delta\gamma$ can be extracted as~\cite{Xu:2017qfs}
\begin{equation} \label{Fcme}
\begin{split}
	f_{\cme} =& \frac{\Delta\gamma_{\cme} \{\spp\}}{\Delta\gamma\{\spp\}}
	= \frac{ A/a - 1 }{ 1/a^{2} - 1 }\,,
\end{split}
\end{equation}
where %$A = \Delta\gamma \{\ssp\} / \Delta\gamma \{\spp\}$,
\begin{equation}\label{eq:A}
    A = \Delta\gamma \{\ssp\} / \Delta\gamma \{\spp\}\,,
\end{equation}
and the geometry fluctuation factor $a$ can be measured by %$a = v_{2} \{\ssp\} / v_{2} \{\spp\}$.
\begin{equation}\label{eq:a}
    a = v_{2} \{\ssp\} / v_{2} \{\spp\}\,.
\end{equation}
The $\fcme$ is effectively determined by the quantity $A/a$, which is the double ratio of $\dg/v_2$ w.r.t.~\ssp\ to that w.r.t.~\spp.

In reality, due to possible complications of \cme\ signal generation and/or evolution with the bulk medium, the \cme\ signal reduction from \ssp\ to \spp\ may not equal to the same factor $a$~\cite{private}. Suppose the reduction factor is $b$, i.e.~$\dg_{\cme}\{\spp\}=b\dg_{\cme}\{\ssp\}$, then it is straightforward to arrive at $f_{\cme}=\frac{A/a-1}{1/ab-1}$ following the algebra in Ref.~\cite{Xu:2017qfs}. The difference would be simply a scaling factor $\frac{1/a^2-1}{1/ab-1}$ once $b$ can be reliably obtained from theory.
We will not dwell on this complication in our present work, and will simply assume $b=a$ in the paper.

Experimentally, the \ssp\ can be assessed by the first-order event plane of spectator neutrons measured in zero-degree calorimeters (\textsc{zdc}), and the \spp\ can be assessed by the second-order event plane (\sep) reconstructed from final-state hadrons~\cite{Zhao:2018blc,Zhao:2020utk}. For simplicity, we will continue to use \ssp\ for the former, but will use \sep\ for the latter to distinguish it from \spp\ as {\em nonflow} implications are different. Their azimuthal angles are $\Psi_\ssp$ and $\Psi_\sep$, respectively, and their measurement inaccuracy are corrected by event-plane resolutions~\cite{Poskanzer:1998yz}. 

We will use an asterisk on a variable to signal it contains nonflow, and reserve the original one to contain only ``true" flow.
To calculate the $v_2^*$ and $\dg^*$ w.r.t.~the \sep, the \poi\ should be excluded from the \sep\ reconstruction to avoid self-correlations. Alternatively one may use the cumulant method~\cite{Poskanzer:1998yz},
\begin{equation}\label{eq:v2}
    v_2^*=v_{2,c}^*=\sqrt{\mean{\cos2(\phi_{\alpha}-\phi_c)}}\,,
\end{equation}
where the $\alpha$ and $c$ particles are from the same phase space or from two sub-events symmetric about  midrapidity having a given pseudorapidity ($\eta$) gap;
and 
 \begin{equation}\label{eq:dg3}
 \begin{split}
   C^{*}_{3} =& \mean{ \cos(\phi_{\alpha}+\phi_{\beta}-2\phi_{c})}\,, \\
   \dg^*=& C^{*}_{3}/v_{2,c}^*\,,
\end{split}
\end{equation}
where the third particle $c$ serves as the \sep\ and its elliptic flow parameter $v_{2,c}^*$ is the \sep\ resolution. (By convention greek subscripts $\alpha, \beta$ are used to stand for \poi, and roman letter $c$ is used for the \sep\ measurement tool.) Again, in Eq.~(\ref{eq:dg3}), the three-particle (3p) correlator $C_3^*$ refers to the $\pos-\pss$ difference.
One way to eliminate self-correlations is to separate the \poi\ and the \sep, or equivalently the \poi\ and $c$, in phase space by applying the sub-event method.
For clarity, we will label these quantities by `$\{\sep\}$' in this paper, even though they may be calculated by multi-particle cumulants.

Accordingly, the quantities $a$, $A$, and $\fcme$ in Eqs.~(\ref{Fcme}), (\ref{eq:A}), and (\ref{eq:a}) will also be tagged by an asterisk because they are, in turn, all affected by nonflow. In fact, a nonzero $\fcme^*$ extracted by the \ssp/\spp\ method, in the absence of real \cme\ signal, is by definition all coming from nonflow. This is the subject of the present paper.

%------------------------------------------------------------------------------------------------%

\section{Nonflow effects} \label{sec:nonflow}

Flow is a global correlation--all particles in an event are correlated because of their correlations to a common symmetry plane (\srp, \ssp, or \spp)~\cite{Voloshin:1994mz}. 
Nonflow, on the other hand, refers to any correlations that are not of an origin of the global, event-wise azimuthal correlations to a common symmetry plane~\cite{Borghini:2000cm,Borghini:2006yk}.
%Experimentally the RP or SP is usually measured by ZDC, $\ssp$. 
Because \textsc{zdc} measures spectator neutrons, 
%$\Psi_\ssp$ is a fairly good estimate of the \ssp. Similarly, 
the $v_2\{\ssp\}$ of midrapidity particles measured w.r.t.~$\Psi_\ssp$ is a good estimate of the ``true" elliptic flow (w.r.t.~\ssp); 
there is little nonflow contamination because spectator neutrons are not dynamic and because of the large $\eta$ gap between midrapidity and the \textsc{zdc}. 

The \spp, on the other hand, is assessed by the \sep\ reconstructed from final-state hadrons. There exist nonflow effects in the reconstructed $\Psi_\sep$. The $v_{2}^{*}\{\sep\}$ measured w.r.t.~$\Psi_\sep$, or similarly by the 2p cumulant of Eq.~(\ref{eq:v2}), is therefore contaminated by nonflow effect. 
The nonflow in $v_2^*$ is mainly from 2p correlations (3p ones are comparatively negligible):
\begin{equation}
    v_{2,{\rm nf}}^2=\mean{\cos 2(\phi_{\alpha} - \phi_c)}_{\rm nf} %=\frac{C_{\Delta\phi} N_{\rm 2p}}{ 2 N (2 N-1)/2}
	=\frac{C_{\Delta\phi} N_{\Delta\phi}}{2N^2}\,,
\end{equation}
where $N_{\Delta\phi}$ is the number of correlated pairs and $C_{\Delta\phi} \equiv \langle \cos 2\Delta\phi \rangle$ is the 2p correlation within the pair.
Note $N_{\Delta\phi}$ includes all ({\em charge-independent}) correlated pairs, not just \pos\ pairs (such as those from resonance decays) but also \pss\ pairs (from jets etc.).
%The nonflow component of $v_2$ is given by $v_2 = \sqrt{\langle \cos 2(\alpha-\beta) \rangle} \approx \sqrt{\frac{C_{\Delta\phi} N_{2p}}{ 2 N^{2} }}$.
Since $N_{\Delta\phi} \propto N$, nonflow decreases with increasing multiplicity.
In general, the $v_2^*\{\sep\}$ from 2p cumulant contains both flow and nonflow: 
\begin{equation}
% 	v_2^*\{\sep\} = \sqrt{v_2^2\{\sep\} + \frac{ C_{\Delta\phi} N_{\rm 2p} }{ 2 N^{2}}}\,.
    v_2^*\{\sep\} = \sqrt{v_2^2\{\sep\} + v_{2,{\rm nf}}^2}\,.
\end{equation}
%where $v_{2,f}$ means the flow contribution.

The major background contribution to $\Delta\gamma$ is the flow-induced background, given by Eq.~(\ref{eq:bkgd}). Let us refer to this contribution as $C_{3,{\rm 2p}}$, standing for 2p contribution to the 3p correlator $C_3$.
In terms of $C_3$, before $v_{2,c}^*=v_2^*$ is divided out, we have
\begin{equation}
	C_{3,{\rm 2p}} %= \langle \cos(\alpha+\beta-2c) \rangle_{\pos - \pss} 
	%\approx \frac{C_{\rm 2p}N_{\rm 2p}}{N^{2}} v_{2,{\rm 2p}} v_{2,{\rm f}} \,,
	= \frac{C_{\rm 2p}N_{\rm 2p}}{N^{2}} v_{2,{\rm 2p}} v_2 \,.
	\label{eq:c3flow}
\end{equation}
%where $N_{\rm 2p}$ is the number of 2p clusters from, for example, resonance decays. [This is already defined for Eq.2]
Note that the correlation between the 2p pair and particle $c$ here is due to pure flow (nonflow effect is discussed below), so what matter in Eq.~(\ref{eq:c3flow}) are their true flows ($v_{2,{\rm 2p}}$ and $v_{2}$, without asterisk marks). Although one element of this background is 2p nonflow correlations as previously discussed, we do not label the l.h.s.~quantity by an asterisk but consider it as flow-induced.
We have taken the shorthand notation for the charge-dependent 2p correlation as
$C_{\rm 2p} = \left\langle \cos(\phi_\alpha+\phi_\beta-2\phi_{\rm 2p}) \right\rangle$.
%Again, we have taken $C_{\rm 2p}$ to refer to the \pos-\pss\ excess.

Note the form of Eq.~(\ref{eq:c3flow}) is the same for both \sep\ and \ssp, so we simply use $v_{2,{\rm 2p}}$ and $v_2$ to refer to those w.r.t.~$\ssp$ or $\sep$.
This background is the main background in the 3p correlator, and is present in both $C_3\{\sep\}$ and $C_3\{\ssp\}$, proportional to the respective $v_2\{\sep\}$ and $v_2\{\ssp\}$. This proportionality, together with the inverse proportionality of the \cme\ effect, renders the validity of the \ssp/\spp\ method as discussed previously.

There is an additional background contribution to $\dg^*\{\sep\}$; this is the charge-dependent (i.e.~between the $\alpha$ and $\beta$ \poi's but irrespective to the charge of particle $c$) 3p nonflow correlations to $C_3^*$, and subsequently propagated to $\Delta\gamma^* = C_{3}^*/v_{2}^*$. It can be expressed as
\begin{equation}\label{eq:c3ep}
	C_{3,{\rm 3p}}^*\{\sep\} %= \langle \cos(\alpha+\beta-2c) \rangle_{\pos-\pss} 
	%\approx \frac{ C_{\rm 3p} N_{\rm 3p} }{ 2 N^{3} }
    = \frac{ C_{\rm 3p} N_{\rm 3p} }{ 2 N^{3} }\,,
\end{equation}
where $N_{\rm 3p}$ is the number of correlated 3p triplets, and 
$C_{\rm 3p} = \mean{\cos(\phi_\alpha + \phi_\beta - 2\phi_c)}_{\rm 3p}$ where the three particles ($\alpha$, $\beta$, $c$) belong to the same triplet. A major source of the charge-dependent 3p nonflow correlations may come from di-jet correlations.
The total 3p correlators are therefore given by
\begin{subequations} \label{eq:C3}
\begin{align}
    C_{3}\{\ssp\} =& \frac{C_{\rm 2p}N_{\rm 2p}}{N^{2}} v_{2,{\rm 2p}}\{\ssp\} v_2\{\ssp\} \,, \label{eq:C3a}\\
    C_{3}^*\{\sep\} =& \frac{C_{\rm 2p}N_{\rm 2p}}{N^{2}} v_{2,{\rm 2p}}\{\sep\} v_2\{\sep\} + \frac{ C_{\rm 3p} N_{\rm 3p} }{ 2 N^{3} }\,. \label{eq:C3b}
\end{align}
\end{subequations}
Note again that we have taken the ``$C$'' quantities to refer to the $\pos-\pss$ differences (i.e.~charge-dependent, where the charge refers to the $\alpha,\beta$ particles).

%------------------------------------------------------------------------------------------------%
%put it here to try to put the figure at bottom of this page so it's close to the text discussing the figure, but it doesn't work.

\begin{figure*}
	\includegraphics[width=0.32\linewidth]{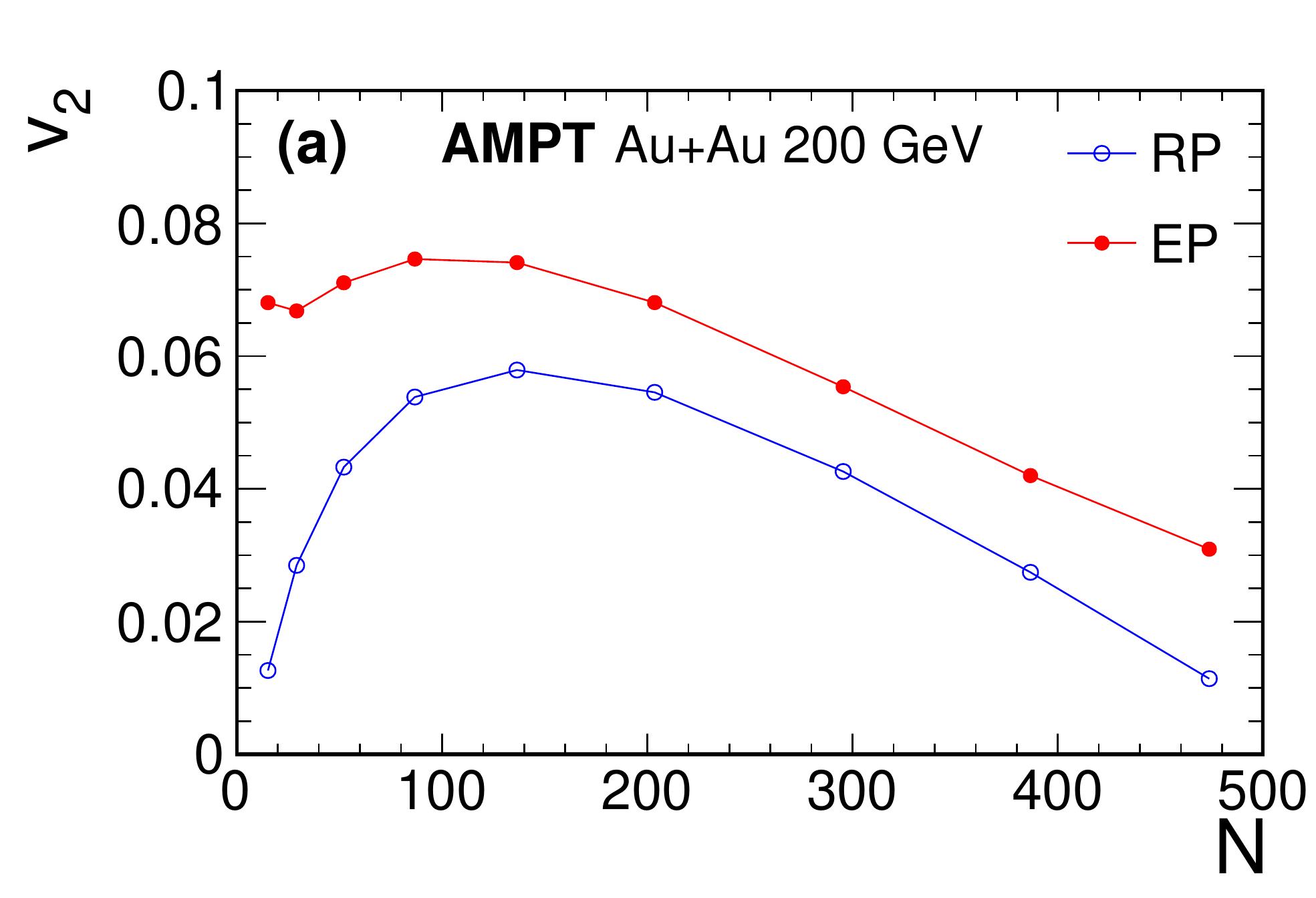}
	\includegraphics[width=0.32\linewidth]{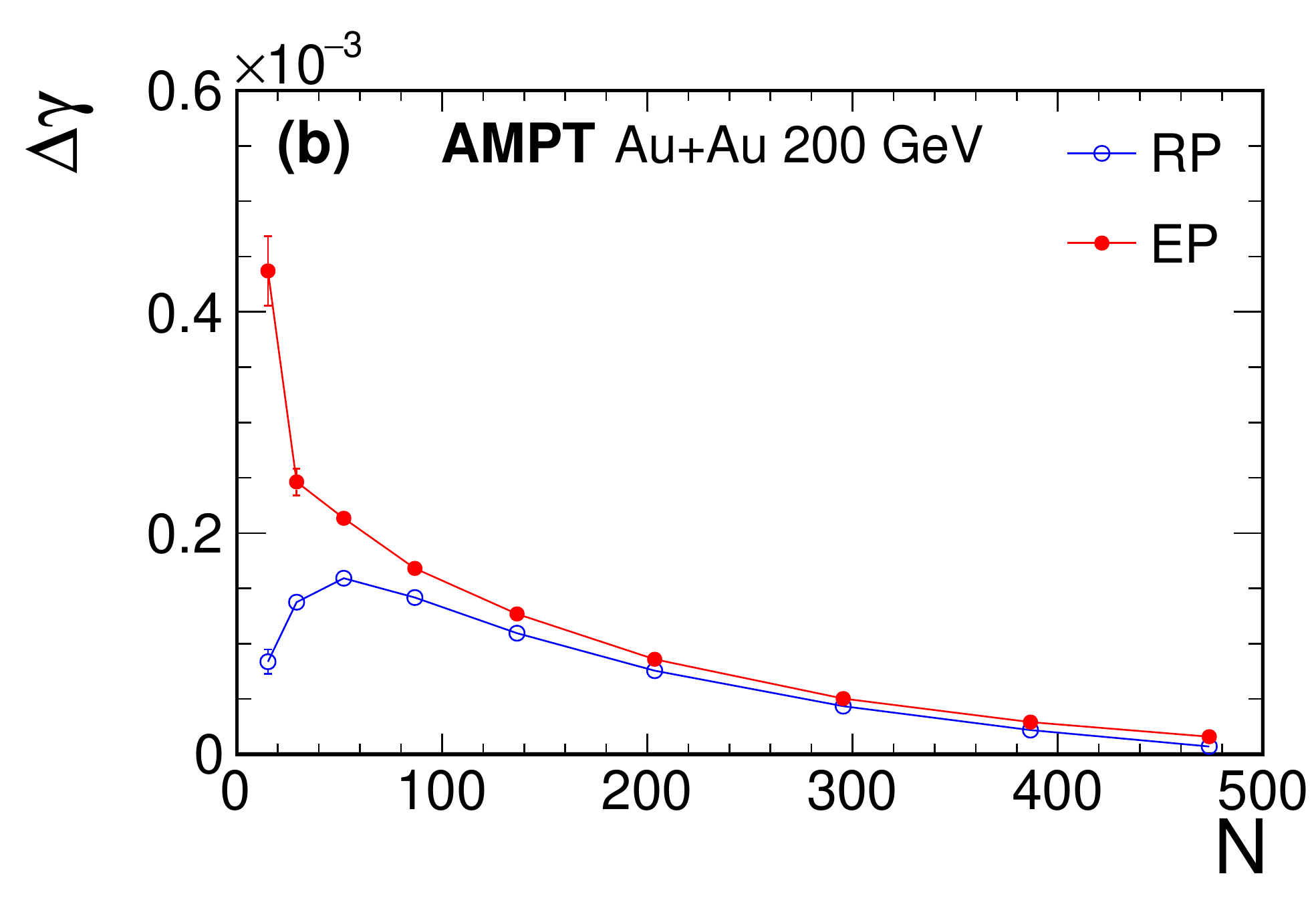}
	\includegraphics[width=0.32\linewidth]{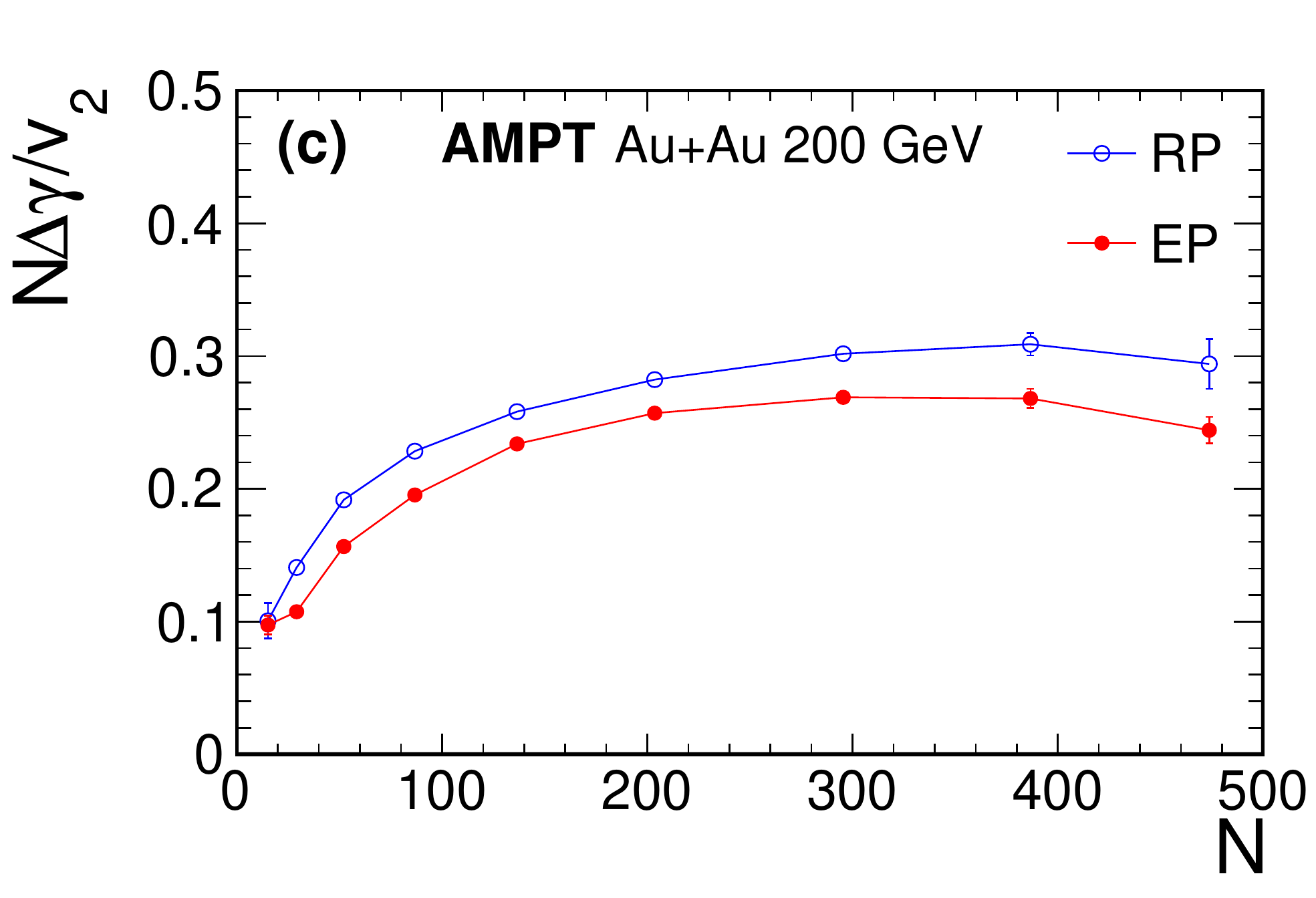}
	\includegraphics[width=0.32\linewidth]{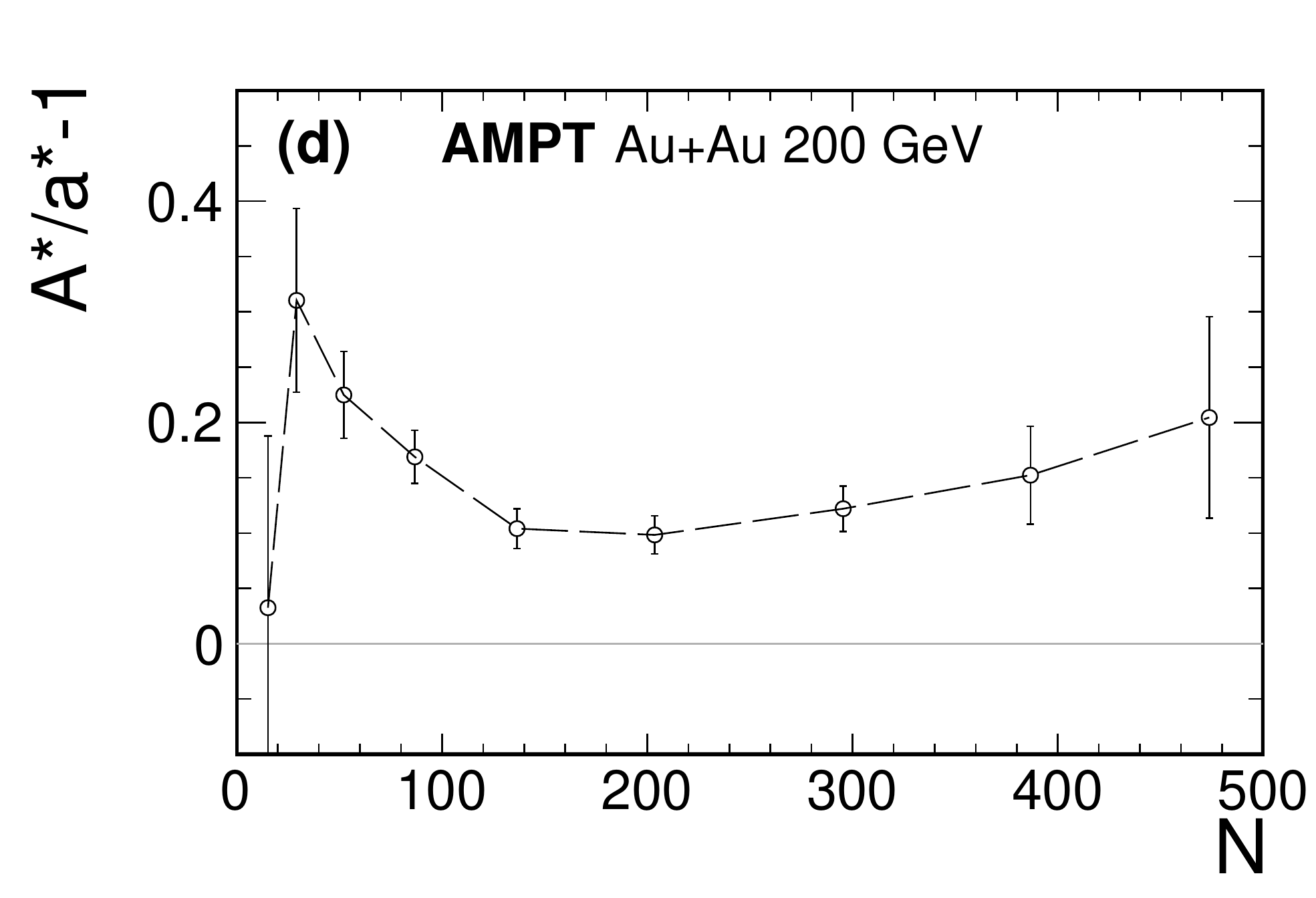}
	\includegraphics[width=0.32\linewidth]{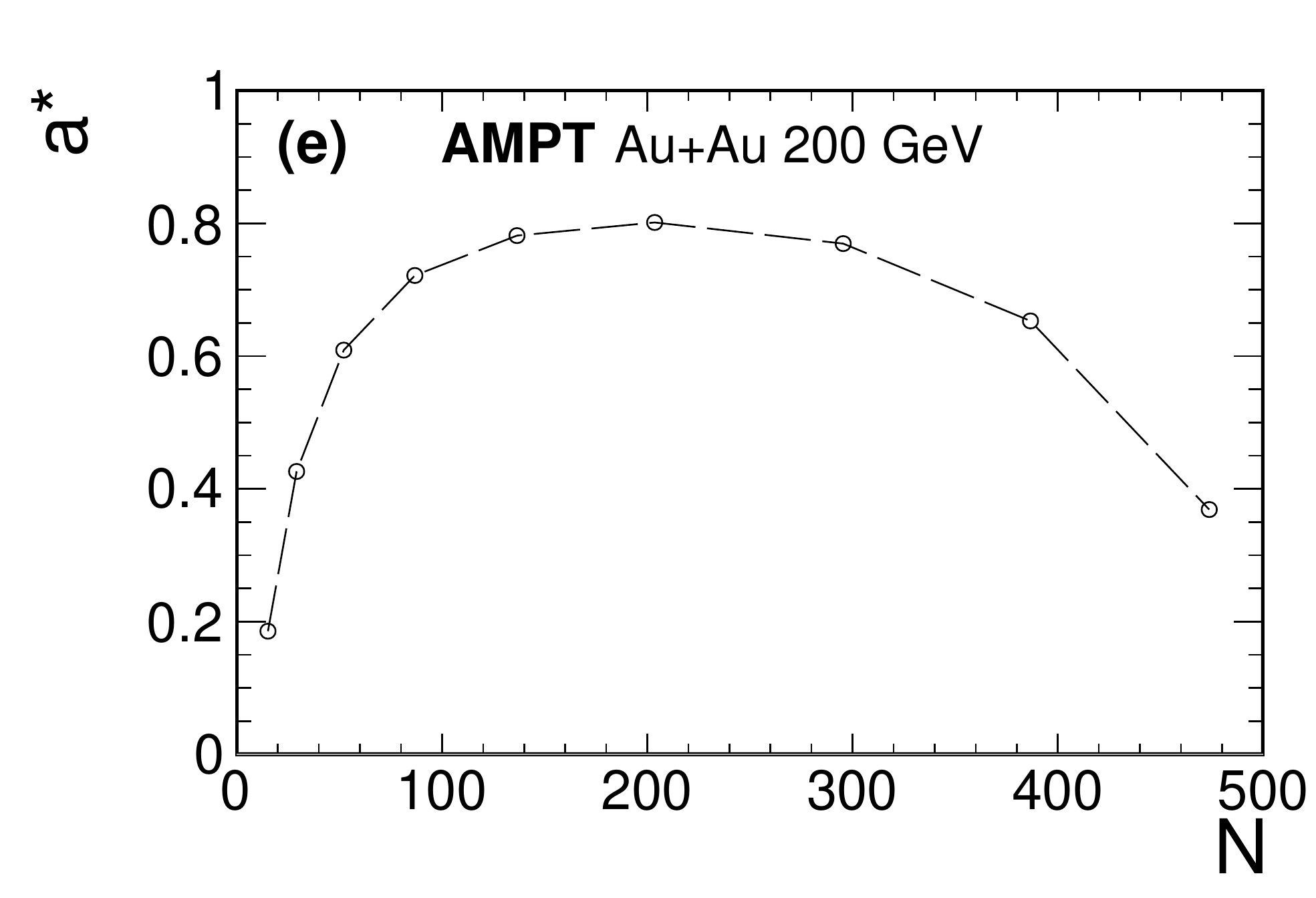}
	\includegraphics[width=0.32\linewidth]{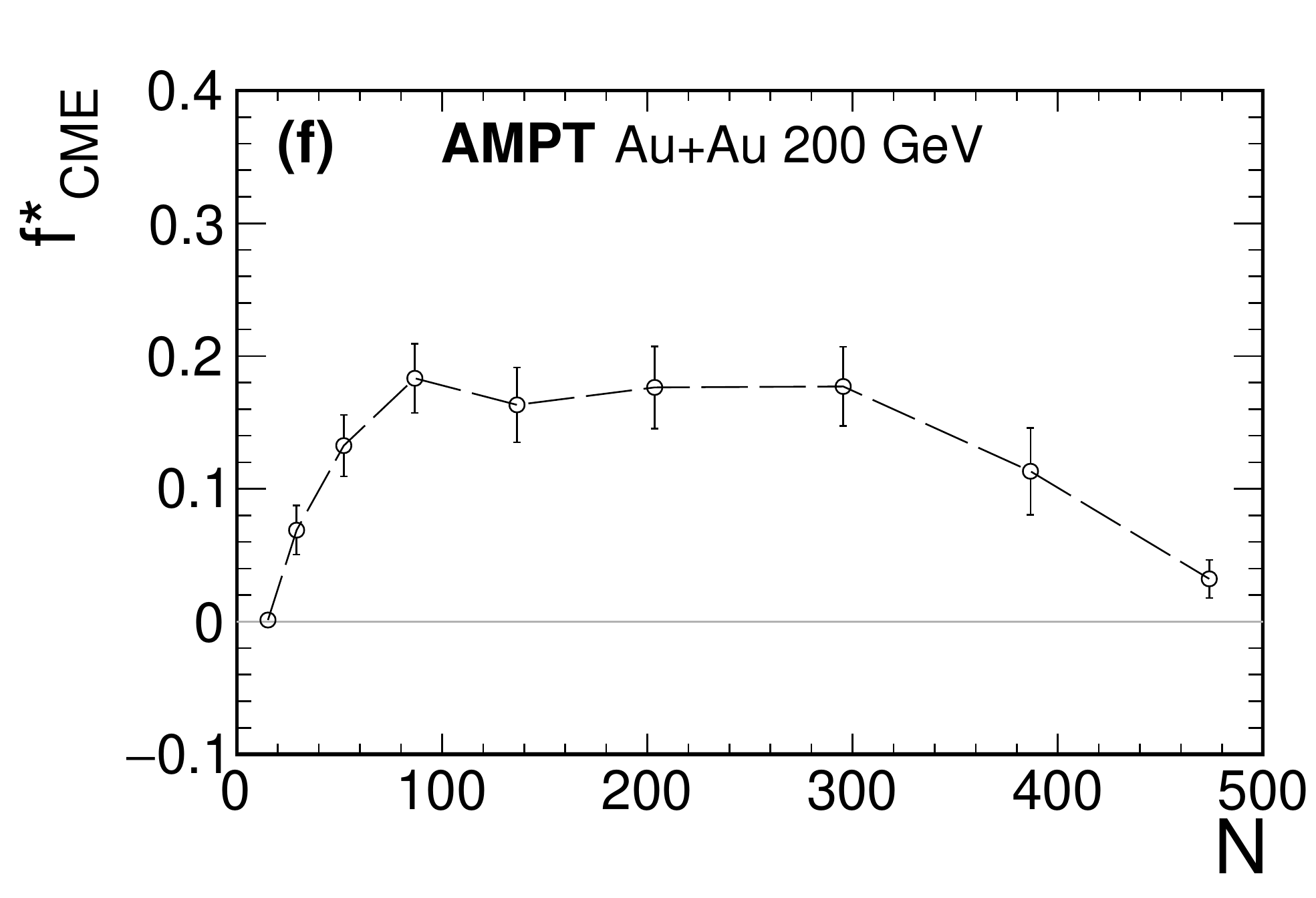}
	\caption{\ampt\ simulation results as functions of $N=(N_++N_-)/2$, the \poi\ single-charge multiplicity, in 200~GeV Au+Au collisions: (a) elliptic flow $v_2$, (b) charge-dependent 3p correlator $\dg$, (c) $N\Delta\gamma/v_{2}$ w.r.t.~\srp\ and \sep\ (the former is referred to as $\epsilon_2^{\ampt}$, see Eqs.~(\ref{eq:bkgd}) and (\ref{eq:e2})), (d) $A^*/a^*-1$ ($\equiv\epsilon_{\ampt}$, which approximately equals to the nonflow contamination $\enf$ in $v_2$, see Eqs.~(\ref{eq:ampt}) and (\ref{eq:Aa})), (e) $a^*$ by Eq.~(\ref{eq:astar}), and (f) the calculated $\fcme^*$ by Eq.~(\ref{Fcme}). The \poi\ and particle $c$ (for \sep) are from $|\eta|<1$ and $0.2<p_T<2$~GeV/$c$. All errors are statistical, with total 377 million \ampt\ mini-bias events.}
	\label{fig:ampt}
\end{figure*}

\section{Simulation Results}\label{sec:model}

% In the pure-background simulations above, the $f_{\text{CME}}$'s deviation from zero 
% can be explained by the different behaviors between 2p and 3p nonflows.

In this section, we use the \ampt\ model~\cite{Zhang:1999bd,Lin:2004en} and the \hijing\ model~\cite{Wang:1991hta,Gyulassy:1994ew} to simulate Au+Au collisions at $\snn = 200 \text{ GeV}$.
No \cme\ signal is included in either model.
The \ampt\ version is v1.25t4cu2/v2.25t4cu2 in which string melting is implemented and the total charge conservation is ensured.
The \hijing\ version is v1.411. %, and charge conservation is ensured in all \hijing\ versions.
In \ampt\ simulation hadronic cascade is included by setting the parameter $\tt{NTMAX=150}$. In \hijing\ simulation jet-quenching is turned on.
About 377 million and 592 million mini-bias events (impact parameter range $0<b<16 \text{ fm}$) are generated by \ampt\ and \hijing, respectively. 
We divide those events into nine centrality bins corresponding to 0-5\%, 5-10\%, 10-20\%, 20-30\%, 30-40\%, 40-50\%, 50-60\%, 60-70\%, and 70-80\% of all events generated, according to the midrapidity charged hadron multiplicity within $|\eta|<0.5$, similar to what is done in the STAR experiment~\cite{Adams:2003xp,Abelev:2008ab}. We will display our results as functions of centrality percentile and $dN/d\eta$, the corresponding charged hadron multiplicity pseudorapidity density, which is approximately the single-charge \poi\ multiplicity ($N$) within $|\eta|<1$.

For \srp, we simply used $\rp=0$ as set in the models. It is also a good estimation of the \ssp, so we will use \srp\ in place of \ssp. 
The \spp\ can be estimated by the \sep\ reconstructed from particles in the same phase space as the \poi, defined to be hadrons within $|\eta| < 1$ and $0.2 < p_{T} < 2.0 \text{ GeV}/c$ (or $0.2 < p_{T} < 1.0 \text{ GeV}/c$) in this study. As aforementioned, the \sep\ method can be replaced by particle cumulant method in calculating $v_2^*$ (and also $C_3^*$) as we adopted in this work. They contain nonflow contributions because of particle correlations. 

Figure~\ref{fig:ampt}(a) shows the $v_2\{\srp\}$ and $v_2^*\{\sep\}$ as functions of $N$ in \ampt. The $v_2^*\{\sep\}$ is significantly larger than the $v_2\{\srp\}$, primarily because of geometry fluctuations (so $v_2\{\sep\}>v_2\{\srp\}$); this difference is exploited in the \ssp/\spp\ method. In addition, there is a relatively minor contribution from nonflow to the difference (i.e.~$v_2^*\{\sep\}>v_2\{\sep\}$); although minor in the difference, it has non-negligible effect on the extracted $\fcme^*$ as we discuss in this paper. 
%In \ampt, $v_{2}$ contains both flow and nonflow. 
Figure~\ref{fig:ampt}(b) shows the $\dg\{\srp\}$ calculated w.r.t.~\srp\ and $\dg^*\{\sep\}$ calculated by the 3p correlator in \ampt.
The $\dg^*\{\sep\}$ is larger than the $\dg\{\srp\}$, primarily because of the correspondingly larger $v_2\{\sep\}$ than $v_2\{\srp\}$ (see Eq.~(\ref{eq:c3flow})).
Since \ampt\ ``destroys" minijets from \hijing\ in its model initialization, the $C_3^*\{\sep\}$ may have little 3p nonflow contributions, so we may assume $C_3^*\{\sep\}\approx C_3\{\sep\}$. Under this assumption, according to Eqs.~(\ref{eq:c3flow}) and (\ref{eq:dg3}), the $\dg\{\sep\}/v_2\{\sep\}$ and $\dg\{\srp\}/v_2\{\srp\}$ would be the same after properly accounting for the respective true flow $v_2$, because presumably
%$\frac{\dg\{\srp\}}{v_2\{\srp\}} = \frac{C_{\rm 2p} N_{\rm 2p} v_{2,{\rm 2p}}\{\srp\} }{ N^2 v_2\{\srp\}}$.
$v_{2,{\rm 2p}}\{\sep\}/v_2\{\sep\} = v_{2,{\rm 2p}}\{\srp\}/v_2\{\srp\}$.
Thus we show in Fig.~\ref{fig:ampt}(c) the ratios of $N\dg\{\srp\}/v_2\{\srp\}$ and $N\dg^*\{\sep\}/v_2^*\{\sep\}$, where $N$ is multiplied to better show the magnitudes.
The former is the following charge-dependent 2p correlation strength (see Eqs.~(\ref{eq:c3flow}) and (\ref{eq:dg3})),
\begin{equation}
    \epsilon_2\equiv\frac{C_{\rm 2p} N_{\rm 2p} v_{2,{\rm 2p}} }{ N v_2}\,;
    \label{eq:e2}
\end{equation}
namely,
\begin{equation}
    \epsilon_2=N\frac{\dg\{\srp\}}{v_2\{\srp\}}=N\frac{\dg\{\sep\}}{v_2\{\sep\}}\,.
\end{equation}
We will refer to this $\epsilon_2$ from \ampt\ as $\epsilon_2^{\ampt}$.
It increases somewhat from peripheral to central collisions. 
The value $\epsilon_2^{\ampt}\sim 0.3$--$0.4$ in mid-central to central collisions makes sense as we roughly expect $C_{\rm 2p} \sim 0.65$, $N_{\rm 2p}/N \sim 0.3$, $v_{2,{\rm 2p}}/v_{2} \sim 2$~\cite{Wang:2016iov}.

The $v_2\{\sep\}$ without nonflow contamination is of course unknown a priori, one measures only the nonflow contaminated $v_2^*\{\sep\}$. 
% Scaling $\dg^*\{\sep\}$ by the ``measured" $v_2^*\{\sep\}$, however, we have
That is, 
\begin{equation}
%\begin{split}
%	\left(\frac{\dg^*\{\sep\}}{v_2^*\{\sep\}}\right)_{\ampt}
	\frac{\dg^*\{\sep\}}{v_2^*\{\sep\}}
	=\frac{C_3\{\sep\}}{v_2^*\{\sep\}^2}
%	=& \frac{C_{2p} N_{2p} v_{2,2p} v_{2,f} }{N^{2} v_{2}^{2}} \frac{\der N_{ch}}{\der \eta}
%	= \frac{C_{2p} N_{2p} v_{2,2p} v_{2,f} }{ N v_{2}^{2}} \\
    % = \frac{C_{\rm 2p} N_{\rm 2p} v_{2,{\rm 2p}}\{\sep\} }{ N^2 v_2\{\sep\}} \frac{1}{1+\enf}\,,
    = \frac{\epsilon_2}{N} \cdot \frac{1}{1+\enf}\,,
%\end{split}
\label{eq:ampt}
\end{equation}
where 
\begin{equation}
    \enf\equiv v_{2,{\rm nf}}^2/v_2^2\,,
\end{equation}
and we have assumed no charge-dependent 3p nonflow contributions in \ampt\ (i.e.~$C_3^*\{\sep\}\approx C_3\{\sep\}$) as aforementioned.
Because of the nonflow in $v_2^*\{\sep\}$, the $\dg^*\{\sep\}/v_2^*\{\sep\}$ is slightly smaller than the $\dg\{\srp\}/v_2\{\srp\}$. % for most of the centralities. 
%(There is perhaps a hint of remaining 3p correlations, due to the incomplete scramble in \ampt\ initialization, manifested in very peripheral collisions.)
In turn, 
the quantity
\begin{equation}
    \frac{A^*}{a^*}=\frac{\dg\{\srp\}/v_2\{\srp\}}{\dg^*\{\sep\}/v_2^*\{\sep\}}
    \label{eq:Aa}
\end{equation} 
is larger than unity; 
let us denote $\epsilon_{\ampt}\equiv A^*/a^*-1$ 
and this is shown in Fig.~\ref{fig:ampt}(d). 
If there is no charge-dependent 3p correlations in \ampt, then $\epsilon_{\ampt}=\enf$ (see Eq.~(\ref{eq:ampt})) would be a good estimate of the nonflow in $v_2^*$ from \ampt;
%The nonflow $\enf$ 
Fig.~\ref{fig:ampt}(d) indicates that it is on the order of 10--20\% depending on centrality. 
This recipe of estimating nonflow by $A^*/a^*-1$ cannot be readily applied to real data because of the potential 3p nonflow contributions to $\dg^*\{\sep\}$ in the real data, which we will discuss later. Of course, if \ampt\ also contains significant charge-dependent 3p nonflow correlations, then the $\enf$ estimate here is also questionable. We will return to this point in Sect.~\ref{sec:realdata}.

From Eq.~(\ref{Fcme}), the larger-than-unity $A^*/a^*$ would result in a positive $\fcme^*=\epsilon_{\ampt}/(1/{a^*}^2-1)$. %$\fcme^*=(A^*/a^*-1)/(1/{a^*}^2-1)$. 
Here the factor $a^*$ is measured by 
\begin{equation}\label{eq:astar}
    a^*=v_2\{\srp\}/v_2^*\{\sep\}=a/\sqrt{1+\enf}\,,
\end{equation}
which is shown in Fig.~\ref{fig:ampt}(e).
The $\fcme^*$ due to nonflow basically equals to $\enf\approx\epsilon_{\ampt}$ multiplied by a factor determined by $a^*$ (or $a$ as the nonflow effect in $a^*$ makes a minor correction).
With the $\enf$ of the order of 10\% and $a\sim 0.8$ in mid-central collisions, a $\fcme^*$ value of the order of 20\% can result, as shown in Fig.~\ref{fig:ampt}(f). This is a significant effect, whereas \ampt\ itself of course does not contain any \cme. It is worthwhile to note, however, that the $\enf$ in \ampt\ shown in Fig.~\ref{fig:ampt}(d) may not be an accurate estimate of nonflow in experimental data, and \ampt\ does not have significant 3p correlations that may be present in real data. Both of these affect the estimate of nonflow contributions to $\fcme^*$; we will return to this in Sect.~\ref{sec:realdata}.

\begin{figure}[h]
	\includegraphics[width=0.8\linewidth]{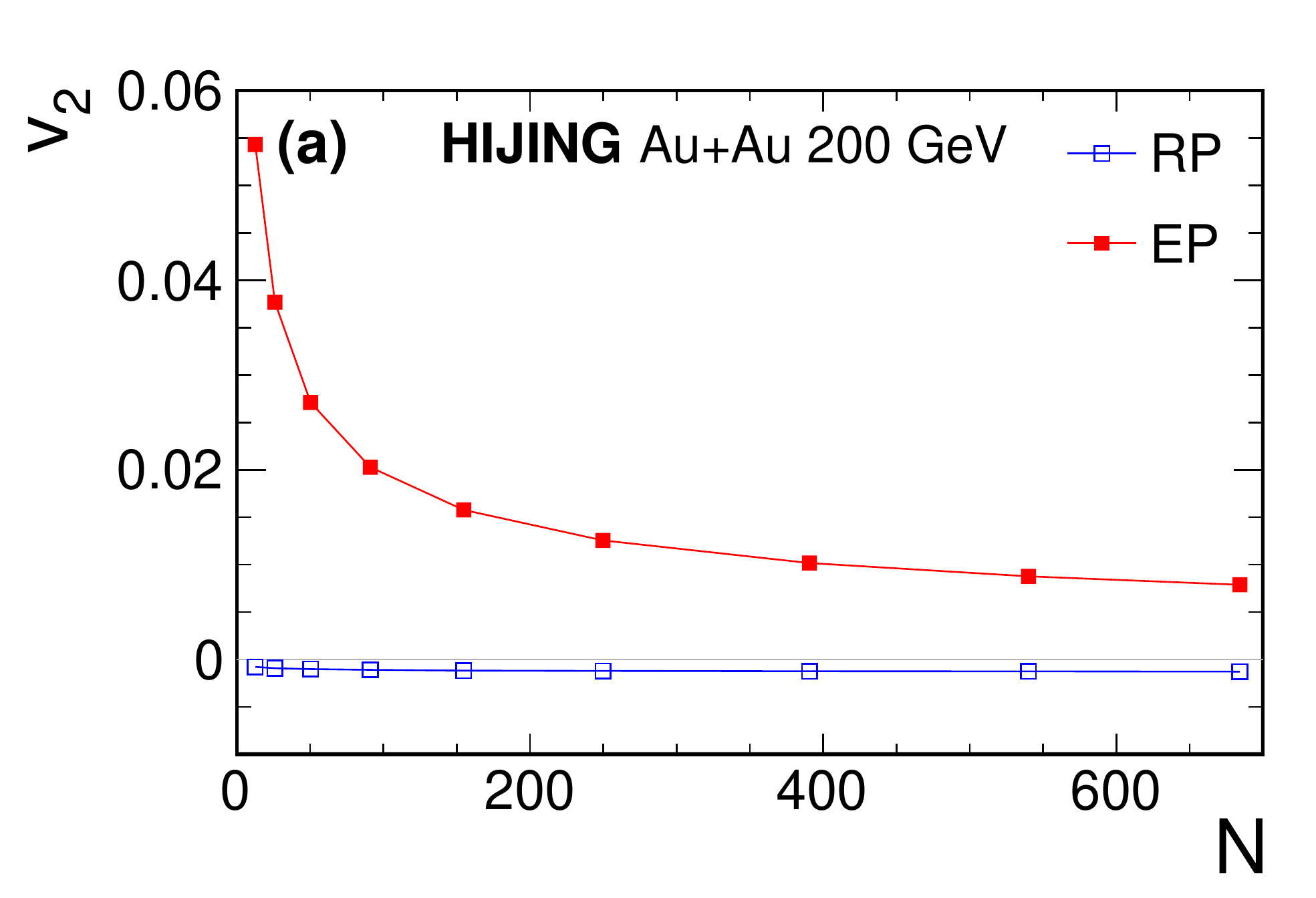}
	\includegraphics[width=0.8\linewidth]{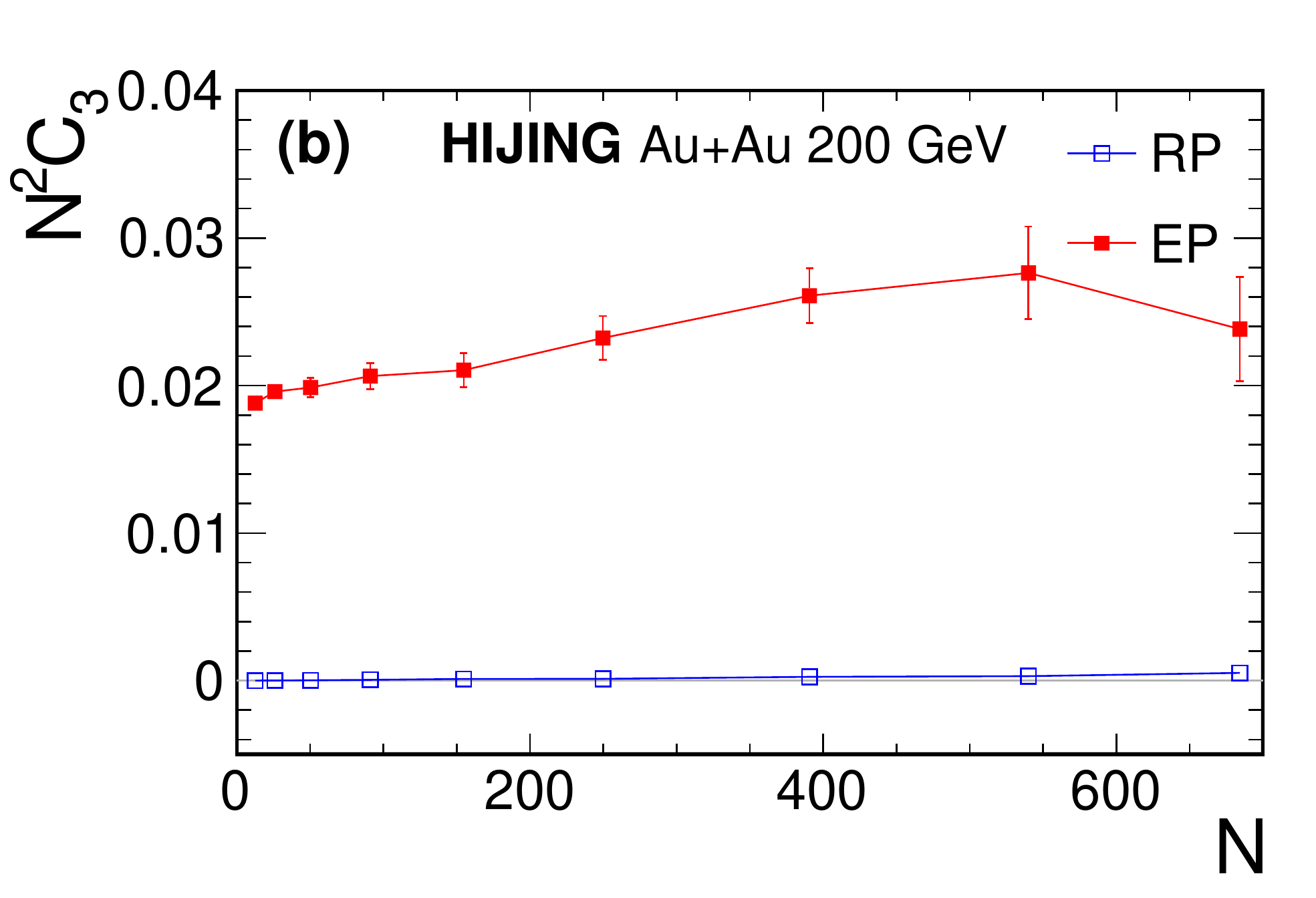}
	\caption{\hijing\ simulation results as functions of $N=(N_++N_-)/2$, the \poi\ single-charge multiplicity, in 200~GeV Au+Au collisions: (a) elliptic anisotropy $v_2$, and (b) charge-dependent 3p correlator $N^2 C_3$ w.r.t.~\srp\ and \sep\ (the latter is referred to as $\epsilon_3=\epsilon_3^{\hijing}$, see Eqs.(\ref{eq:c3ep}), (\ref{eq:C3b}), and (\ref{eq:e3})). The \poi\ and particle $c$ are from $|\eta|<1$ and $0.2<p_T<2$~GeV/$c$. All errors are statistical, with 592 million \hijing\ mini-bias events.}
	\label{fig:hijing}
\end{figure}

Let us now turn to \hijing.
Figure~\ref{fig:hijing}(a) shows the $v_2\{\srp\}$ and $v_2^*\{\sep\}$ from \hijing\ as functions of $N$. The small but negative $v_2\{\srp\}$ is a result of jet-quenching--more particles are generated perpendicular to the \srp\ because of the longer pathlengths jets traverse. Such an azimuthal modulation is global and technically has no distinction from ``real" flow, so we will just refer to it in this paper as ``true" $v_2$. 
Figure~\ref{fig:hijing}(a) shows that the $v_2^*\{\sep\}$ is significantly larger and positive. The major contribution 
to $v_2^*\{\sep\}$ in \hijing\ is nonflow; the jet-quenching induced anisotropy is negligible in $v_2^*\{\sep\}$.
Thus, the flow-induced background $\dg_{\rm bkgd}$ of Eq.~(\ref{eq:bkgd}) or Eq.~(\ref{eq:c3flow}) is small in \hijing; the $\dg^*\{\sep\}$ in Eq.~(\ref{eq:C3b}) will be dominated by the second term, 3p nonflow correlations. In a previous work~\cite{Zhao:2019kyk}, we have shown that $N\dg^*\{\sep\}/v_2^*\{\sep\}$ in \hijing\ is large and has a weak centrality dependence, indicating a good degree of factorization of the 3p (such as di-jet) correlations into 2p correlations.
In Fig.~\ref{fig:hijing}(b) we show directly the 3p correlators, %given by Eq.~(\ref{eq:c3flow})
multiplied by $N^2$, namely $N^2C_3^*\{\sep\}$ and $N^2C_3\{\srp\}$. It is shown, indeed, that $C_3^*\{\sep\}$ is significantly larger than $C_3\{\srp\}$, with the latter being negligible. This indicates that the 3p nonflow contribution dominates in $C_3^*\{\sep\}$ over the ``flow"-induced contribution in \hijing. The $N^2C_3^*\{\sep\}$ in Fig.~\ref{fig:hijing}(b), which we will refer to as $\epsilon_3^{\hijing}$, essentially gives the charge-dependent 3p correlation strength (see Eq.~(\ref{eq:c3ep})),
\begin{equation}\label{eq:e3}
    \epsilon_3\equiv\frac{C_{\rm 3p} N_{\rm 3p}}{2N}\,.
\end{equation}
Its strength has only modest increase with centrality in \hijing.

Unlike \ampt, \hijing\ does not have a significant flow-induced background, so it is not meaningful to extract a $\fcme^*$ value from \hijing\ like we did for \ampt. However, the 3p nonflow correlations in \hijing, that \ampt\ lacks, are useful knowledge to assess additional nonflow effect in a real data analysis which we now attend to.

\section{Implications to real data}\label{sec:realdata}
Real experimental data are probably similar to \ampt\ in terms of flow, and likely contain 3p correlations similar to \hijing.
According to Eq.~(\ref{eq:C3b}) we can write
\begin{equation}
% 	\frac{\dg\{\sep\}}{v_2\{\sep\}} = \frac{ C_{\rm 2p} N_{\rm 2p} v_{2,{\rm 2p}} v_{2,{\rm f}} / N^{2} + C_{\rm 3p} N_{\rm 3p} / (2 N^{3}) }  	{ v_{2,{\rm f}}^{2} + C_{\Delta\phi} N_{\rm 2p} / (2 N^{2})} \,.
	\frac{\dg^*\{\sep\}}{v_2^*\{\sep\}} = 
% 	\frac{ C_{\rm 2p} N_{\rm 2p} v_{2,{\rm 2p}}\{\sep\} v_2\{\sep\} / N^{2} + C_{\rm 3p} N_{\rm 3p} / (2 N^{3})} {v_2^2\{\sep\}(1+\enf)}\,.
% 	\frac{
% 	\frac{C_{\rm 2p}N_{\rm 2p}}{N^{2}}v_{2,{\rm 2p}}\{\sep\} v_2\{\sep\} 
% 	\left(1+ \frac{\epsilon_3/\epsilon_2}{N v_2^2\{\sep\}}\right)
% 	}{v_2^2\{\sep\}(1+\enf)}\,,
% 	\frac{C_{\rm 2p}N_{\rm 2p}v_{2,{\rm 2p}}\{\sep\}}{N^{2}v_{2}\{\sep\}} \cdot
% 	\frac{1+ \frac{\epsilon_3/\epsilon_2}{N v_2^2\{\sep\}}}{1+\enf}\,,
	\frac{\epsilon_2}{N} \cdot
	\frac{1+ \frac{\epsilon_3/\epsilon_2}{N v_2^2\{\sep\}}}{1+\enf}\,,
\end{equation}
% where
% \begin{equation}
%     \epsilon_3\equiv\frac{C_{\rm 3p} N_{\rm 3p}}{2N}\,.
% \end{equation}
where $\frac{\epsilon_3/\epsilon_2}{N v_2^2\{\sep\}}=\frac{C^*_{3,3p}}{C_{3,2p}}$ (see Eqs.~(\ref{eq:c3flow}) and (\ref{eq:c3ep})) is just the relative 3p over 2p contributions to the 3p correlator.
Measurements w.r.t.~\srp\ are not affected by nonflow, so we simply have $\dg\{\srp\}/v_2\{\srp\} = \epsilon_2/N$. From Eq.~(\ref{Fcme}), we obtain
\begin{subequations}\label{eq:fcmedata}
\begin{align}
    \fcme^*=&\left.\left(\frac{1+\enf}{1+\frac{\epsilon_3/\epsilon_2}{N v_2^2\{\sep\}}}-1\right)\right/\left(\frac{1+\enf}{a^2}-1\right)\label{eq:fcmedataA}\\
    =&\left.\left(\frac{1+\enf}{1+\frac{(1+\enf)\epsilon_3/\epsilon_2}{N {v_2^*}^2\{\sep\}}}-1\right)\right/\left(\frac{1}{{a^*}^2}-1\right)\,.\label{eq:fcmedataB}
\end{align}
\end{subequations}
The 2p nonflow effect, $\enf$, increases 2p cumulant $v_2^*\{\sep\}$, and consequently introduces a positive $\fcme^*$ (as in \ampt). The 3p nonflow effect, $\epsilon_3$, increases $C_3^*\{\sep\}$ and $\dg^*\{\sep\}$, and consequently introduces a negative $\fcme^*$. That the two nonflow effects cancel each other to some degree is a neat feature, making the $\fcme^*$ from the \srp/\spp\ method less vulnerable to nonflow. The quantitative conclusion depends of course on the relative magnitudes of the nonflow effects from 2p and 3p correlations.

It is worthwhile to note that, because $\epsilon_3 \ll \epsilon_2$ (by comparing Fig.~\ref{fig:ampt}(c) and Fig.~\ref{fig:hijing}(b)) and normally $Nv_2^2\sim\mathcal{O}(1)$, the flow-induced background (due to charge-dependent 2p correlations) is the leading order term in $C_3^*$ (and $\dg^*\{\sep\}$) and the charge-dependent 3p nonflow correlations are the next-to-leading order (NLO) perturbation; meanwhile the NLO perturbation in $v_2^*$ is the nonflow from charge-independent 2p correlations and the charge-independent 3p nonflow correlations can be neglected.
To the order of the respective NLO terms of
%Treating the nonflow effects to be small in 
$v_2^*$ and $C_3^*$, we may write
\begin{subequations}\label{eq:fcmedataapp}
\begin{align}
    \fcme^*\approx&\left(\enf - \frac{\epsilon_3/\epsilon_2}{N v_2^2\{\sep\}}\right)\left/\left(\frac{1+\enf}{a^2}-1\right)\right.\label{eq:fcmedataappA}\\
    =& \left(\enf - \frac{(1+\enf)\epsilon_3/\epsilon_2}{N {v_2^*}^2\{\sep\}}\right)\left/\left(\frac{1}{{a^*}^2}-1\right)\right.\,.\label{eq:fcmedataappB}
\end{align}
\end{subequations}
However, since nonflow $\enf$ and $(\epsilon_3/\epsilon_2)/(Nv_2^2)$ may not always be small compared to unity (e.g.,~in peripheral collisions), we will nonetheless use Eq.~(\ref{eq:fcmedata}) to estimate the magnitudes of the nonflow effects in $\fcme^*$.

\begin{figure}
    \centering
    \includegraphics[width=1.0\linewidth]{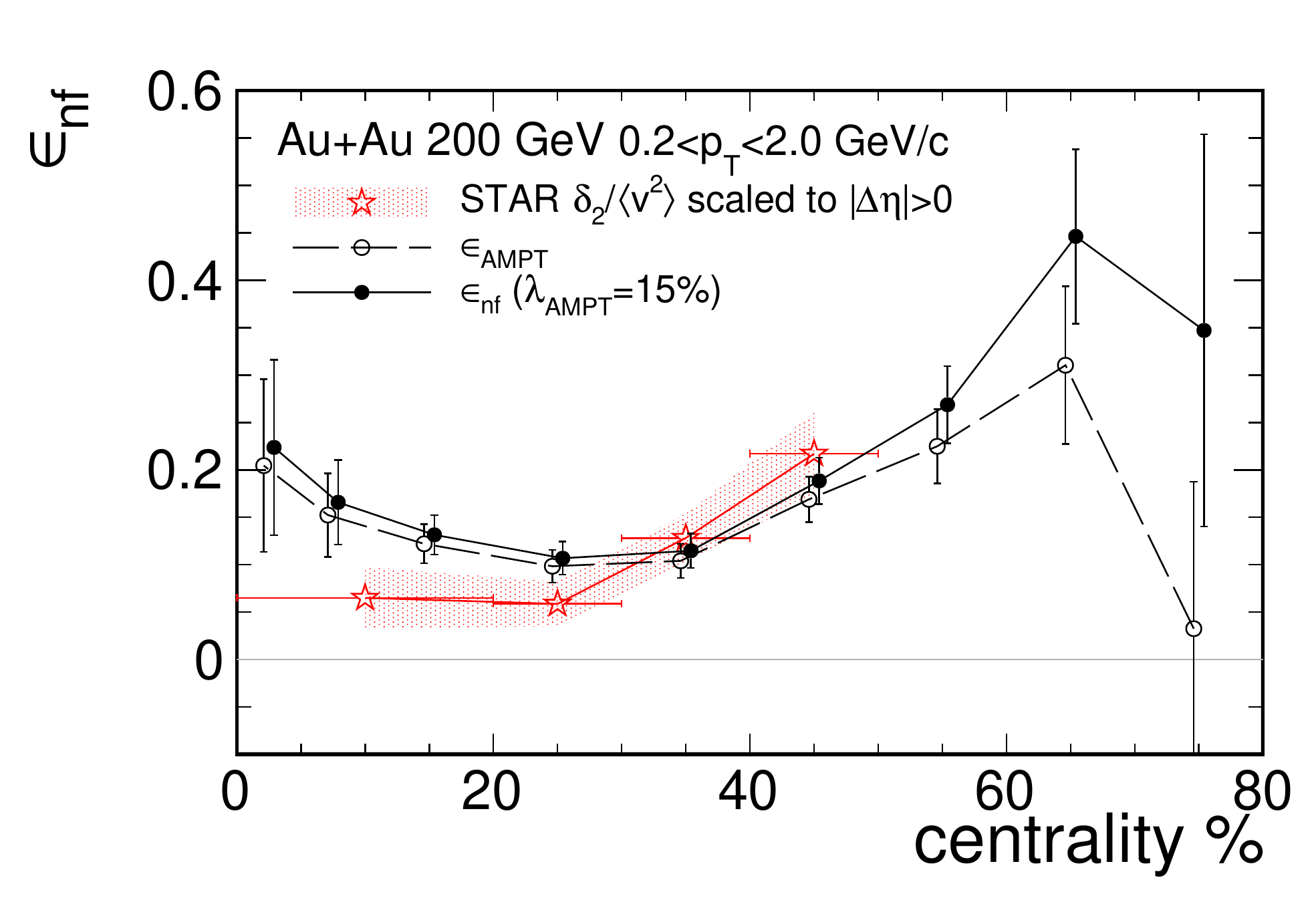}
    \caption{Estimated $v_2$ nonflow as functions of centrality in 200 GeV Au+Au collisions (data points slightly shifted in horizontal axis for clarity). The open (filled) circles are from \ampt, assuming 0\% (15\%) charge-dependent 3p contributions to $\dg^*\{\sep\}$. Errors are statistical. The red stars are STAR data~\cite{Abdelwahab:2014sge}, where the systematic uncertainties are $\pm50\%$ for centrality 0--20\%, $\pm40\%$ for 20--30\%, and $\pm20\%$ for 30--50\%.}
    \label{fig:FitScl}
\end{figure}

\begin{figure}
	\includegraphics[width=1.0\linewidth]{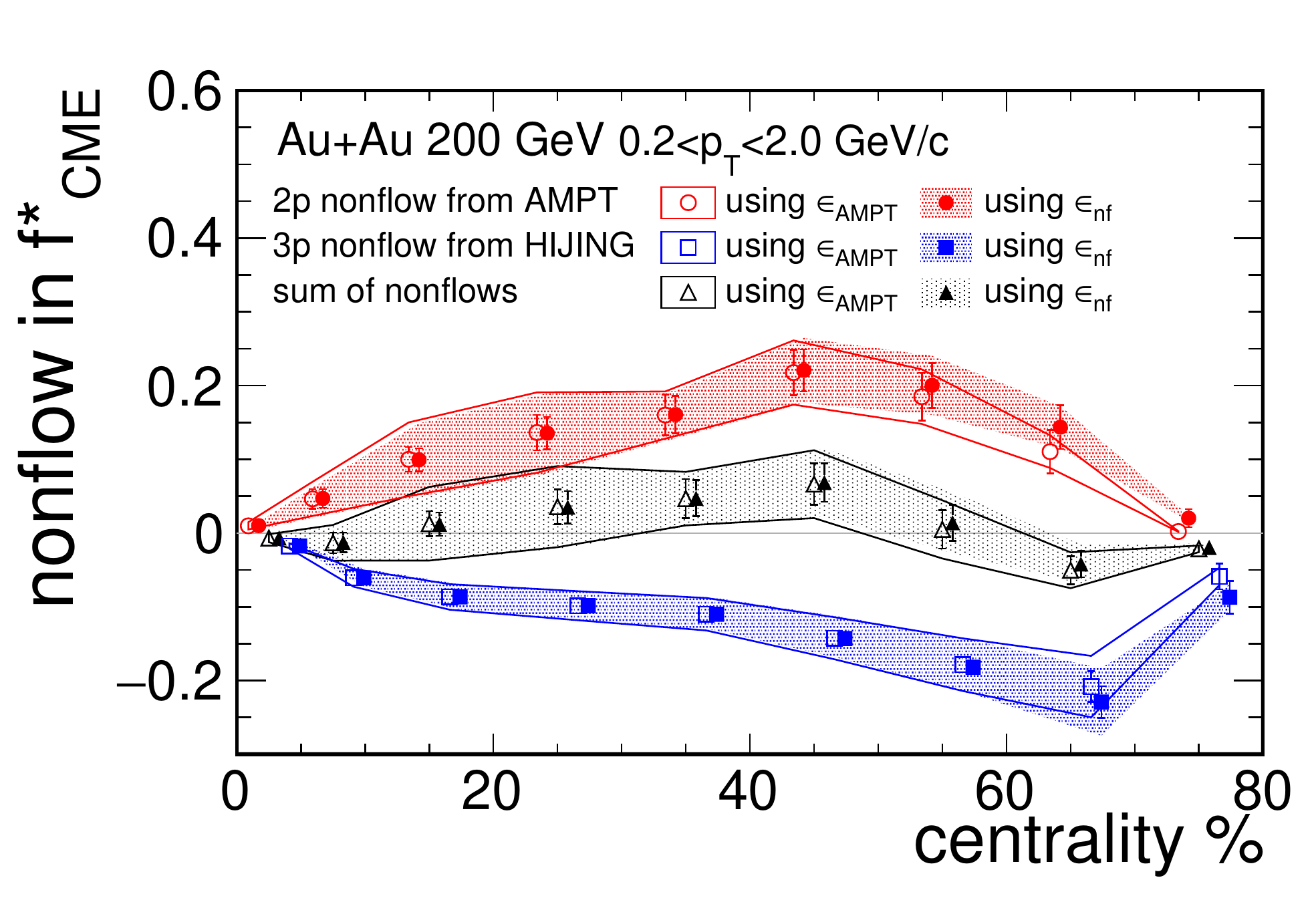}
	\caption{The 2p and 3p nonflow contributions (by Eq.~(\ref{eq:fcmedataapp})) and their net contribution (Eq.~(\ref{eq:fcmedata})) to $\fcme^*$ as functions of centrality in 200~GeV Au+Au collisions, with elliptic flow and geometry fluctuation effect (i.e.~$a^*$) taken from \ampt, and with various assumptions of nonflow contributions. Open markers: charge-independent 2p nonflow as in \ampt, $\enf=\epsilon_{\ampt}$ as from Fig.~\ref{fig:ampt}(d) (i.e.~3p contribution $\lambda_\ampt=0$ in Eq.~(\ref{eq:lambda})) and scaled to experimental measurement~\cite{Abdelwahab:2014sge} (open circles), charge-dependent 3p nonflow correlations as in \hijing, $\epsilon_3=\epsilon_3^{\hijing}\pm20\%$ (open squares), and the sum of the two (open triangles); the open bands are the respective systematic uncertainties in matching to experimental data. Solid markers: corresponding nonflow effects as open markers but with $\lambda_\ampt=15\%$ in Eq.~(\ref{eq:lambda}), the charge-dependent 3p correlations in \ampt\ relative to those in \hijing; the shaded bands are the corresponding systematic uncertaities. The \poi\ and particle $c$ are from $|\eta|<1$ and $0.2<p_T<2$~GeV/$c$. Data points are slightly shifted in horizontal axis for clarity.} 
	\label{fig:fcmeampt}
\end{figure}

The estimate of the 2p nonflow effect boils down to the estimate of $\enf$. 
If $\enf$ is as given by \ampt\ (i.e.~$\enf=\epsilon_{\ampt}$ as in Fig.~\ref{fig:ampt}(d)), then its effect on $\fcme^*$ would be that shown in Fig.~\ref{fig:ampt}(f).
Nonflow has been extensively studied in real data. A data-driven way to estimate nonflow contribution is performed by STAR~\cite{Abdelwahab:2014sge}. We show in Fig.~\ref{fig:FitScl} the estimated nonflow $\enf^{\rm exp}$ in Au+Au collisions for $|\Delta\eta|>0.7$ scaled up by a factor of 2.0 to represent the nonflow contributions without any $\eta$ gap~\cite{Abdelwahab:2014sge}. The systematic uncertainties on the data vary between 20--50\%. 
Also shown in Fig.~\ref{fig:FitScl} by the open circles are the $\epsilon_{\ampt}$ from \ampt\ (i.e.~the data points in Fig.~\ref{fig:ampt}(d)). 
These would be the nonflow $\enf$ in \ampt\ if charge-dependent 3p correlations to $\dg^*\{\sep\}$ can be neglected.
(The filled circles represent those with 3p correlations considered, as explained later in the text.)
The nonflow contribution depends on centrality. 
In 20-30\% or more central collisions, \ampt\ somewhat overestimates the data. In more peripheral collisions, \ampt\ seems to underestimate the data. 
Measurements are unavailable for the peripheral 50-80\% centralities. We extrapolate to those peripheral centralities by fitting the ratio of data over \ampt\ with a linear dependence: $\enf^{\rm exp}/\epsilon_{\ampt}=\fitdataeamptratio$, where the centrality ``cent'' is a number between 0 and 1.
We scale the \ampt\ nonflow $\epsilon_\ampt$ (assuming zero 3p contribution) shown by the open circles in Fig.~\ref{fig:FitScl} by the fitted linear function. 
We use the scaled $\enf$ value to evaluate its contribution to $\fcme^*$ by the first term of Eq.~(\ref{eq:fcmedataappB}), %$\senf\epsilon_{\ampt}/(\frac{1}{{a^*}^2}-1)$, 
$\enf/(\frac{1}{{a^*}^2}-1)$, where $a^*$ is taken from Fig.~\ref{fig:ampt}(e). 
The result is shown in Fig.~\ref{fig:fcmeampt} as function of centrality by the open circles where the statistical errors are from the \ampt\ simulation data sample and 
the open band embraces the experimental uncertainty on nonflow of $\pm$20--50\% (for the extrapolated peripheral range, we assume the same systematic uncertainty of $\pm20$\% as that in the 20--30\% centrality bin).

It is worthwhile to note that here we have effectively used experimental nonflow results in the estimate of $\fcme^*$, by scaling $\epsilon_{\ampt}$ to $\enf^{\rm exp}$. Using \ampt\ as a stepping stone seems unnecessary except the extrapolation to peripheral collisions. However, we will also investigate nonflow effects in other kinematic regions later in the article where   experimental data on nonflow are not readily available. There, we will need to use \ampt\ simulation results and scale them by assuming the same scaling factor as function of centrality parameterized here.

The 3p nonflow effect in $\fcme^*$ can be estimated as follows. 
The $\epsilon_3$ can be obtained from \hijing\ in Fig.~\ref{fig:hijing}(b), $\epsilon_3^{\hijing}=N^2 C_3^*\{\sep\}$, because it has been observed to give a fair description of the small-system collision data at RHIC within 20\%~\cite{Zhao:2019kyk}; so we take $\epsilon_3=\epsilon_3^{\hijing}\pm20\%$.
The flow and flow fluctuation related quantities $N {v_2^*}^2\{\sep\}$ and $a^*$ can be taken from \ampt. 
The $\epsilon_2$ can also be taken from \ampt\ in Fig.~\ref{fig:ampt}(c), but it has been observed that \ampt\ can reproduce only about 60\% of the $\dg^*/v_2^*$ in real data~\cite{Zhao:2019kyk}; so we take $\epsilon_2=1.7\epsilon_2^{\ampt}$. 
(Note that this underestimate of $\epsilon_2$ by \ampt\ does not directly affect the first term of Eq.~\ref{eq:fcmedata}, because it appears in both $\dg^*\{\sep\}$ and $\dg\{\srp\}$ and is cancelled. The $\epsilon_2$ (charge-dependent 2p nonflow) does contribute, in part, to the $\enf$ (charge-independent 2p nonflow), but there are many other charge-independent contributions (e.g.~like-sign particle correlations) that apparently have resulted in an already overestimated $\enf\sim 10$--$20\%$ in \ampt\ for mid-central collisions, as aforementioned.)
The estimated 3p nonflow effect by the second term of Eq.~(\ref{eq:fcmedataappB}),$-\frac{(1+\enf)\epsilon_3/\epsilon_2}{N {v_2^*}^2\{\sep\}}/(\frac{1}{{a^*}^2}-1)$, is shown in Fig.~\ref{fig:fcmeampt} in the open squares with small statistical error bars from the large \hijing\ simulation data sample. 
The open band indicates the $\pm 20\%$ systematic uncertainty, the level of agreement of \hijing\ in describing experimental data. 

The combination of the two, as given by Eq.~(\ref{eq:fcmedata}), would indicate the error one makes in $\fcme^*$ extracted from ``experimental" data, if the 2p and 3p nonflow effects are as given by \ampt\ ($\enf=\epsilon_{\ampt}$, scaled to $\enf^{\rm exp}$) %in 20-30\% centrality) 
and \hijing\ ($\epsilon_3=\epsilon_3^{\hijing}\pm20\%$), respectively. This is shown as the open triangles in Fig.~\ref{fig:fcmeampt}. The accompanying open band is the quadratic sum of the systematic uncertainty estimates on $\enf$ and $\epsilon_3$. 
Note that the net result is not a simple sum of the individual 2p and 3p nonflow effects estimated above via the approximated Eq.~\ref{eq:fcmedataappB}, which is only valid when both effects are small.
As shown by Fig.~\ref{fig:fcmeampt}, nonflow correlations could contribute an artificial $\fcme^*$ signal up to a few percent (in both positive and negative directions), depending on centrality, in Au+Au collisions at 200~GeV. 

In estimating nonflow effects in $\fcme^*$ by Eqs.~(\ref{eq:fcmedata}) and (\ref{eq:fcmedataapp}), we have used \ampt\ to estimate the nonflow effect $\enf$ in $v_2$ and \hijing\ to estimate the 3p nonflow effect $\epsilon_3$ in $\dg^*\{\sep\}$.
We have so far neglected 3p correlations in \ampt\ and attributed the $\dg^*$ (and $\fcme^*$) in \ampt\ all to 2p nonflow, so that $\enf=\epsilon_{\ampt}$. However, \ampt\ does contain some 3p correlations, approximately $\lambda_\ampt=15\%$ of those from \hijing\ as shown by the small-system simulations in Ref.~\cite{Zhao:2019kyk}, presumably due to an incomplete destruction of minijet correlations in \ampt\ model initialization. Since these 3p correlations contribute a negative magnitude to $\fcme^*$, the 2p nonflow effect $\enf$ would be larger than the $\epsilon_{\ampt}$ depicted in Fig.~\ref{fig:ampt}(f). In other words, following Eq.~(\ref{eq:fcmedata}), 
\begin{equation}\label{eq:lambda}
    \enf=\epsilon_{\ampt}+\lambda_\ampt\frac{(1+\enf)\epsilon_3/\epsilon_2}{N {v_2^*}^2\{\sep\}}\,,
\end{equation}
from which we can deduce a new $\enf$ in \ampt.
This is shown by the filled circles in Fig.~\ref{fig:FitScl}.
(Note that the $\lambda_\ampt=15\%$ residual 3p correlation in \ampt\ is only used to calculate an improved $\enf$ by Eq.~(\ref{eq:lambda}); it is not used for any estimate of the 3p correlation contribution to $\dg^*\{\sep\}$, which is obtained from \hijing\ in our study.)
We again fit a linear function to the ratio of data over \ampt\ in Fig.~\ref{fig:FitScl}, $\enf^{\rm exp}/\enf(\lambda_{\ampt}$=$15\%)=\fitdataenfratio$, and then scale the $\enf(\lambda_{\ampt}$=$15\%)$ by the fitted function.
The resultant $\fcme^*$ by Eqs.~(\ref{eq:fcmedataappB}) and (\ref{eq:fcmedataappB}) %using the new $\enf$ 
are depicted in Fig.~\ref{fig:fcmeampt} as the solid markers and shaded bands; note that $\enf$ affects numerically both 2p and 3p nonflow terms.
As seen from Fig.~\ref{fig:fcmeampt}, the end results are not much affected by $\lambda_{\ampt}$; this is because 
the experimental $\enf^{\rm exp}$ is effectively used in the estimate. 
Averaging over 20--50\% centrality, the effect is approximately $\fcme^*=(4\pm5)\%$. %$\fcme^*=\fullnumber$. 
This is for the case where both the \poi\ and particle $c$ (or \sep) are from $|\eta|<1$ and $0.2<p_T<2$~GeV/$c$ (referred to as the full-event method).

\begin{figure}
	\includegraphics[width=1.0\linewidth]{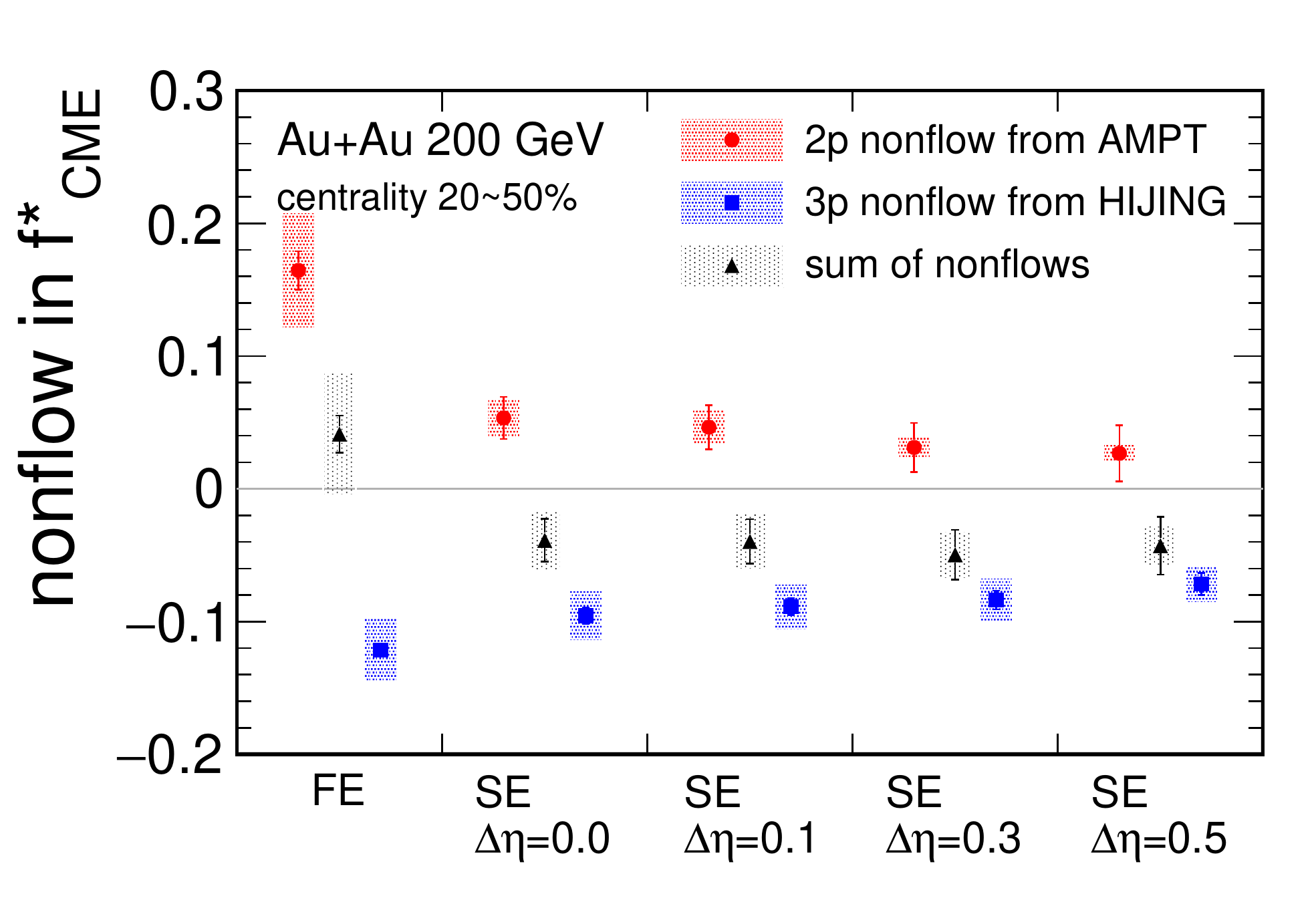}
	\caption{Same as the solid markers in Fig.~\ref{fig:fcmeampt}, but showing the average $\fcme^*$ within 20--50\% centrality in 200~GeV Au+Au collisions, obtained from the full-event (FE) method (i.e.~those in Fig.~\ref{fig:fcmeampt}) along with those from the sub-event (SE) method with various $\eta$ gaps. The \poi\ and particle $c$ are from $|\eta|<1$ and  $0.2<p_T<2$~GeV/$c$.}
	\label{fig:fcme}
\end{figure}

One can reduce nonflow effects by applying an $\eta$ gap between \poi\ and particle $c$, or simply using the sub-event method where the \poi\ are from one sub-event and $c$ from the other. Figure~\ref{fig:fcme} shows the average $\fcme^*$ within the 20--50\% centrality range from the  sub-event method with various $\eta$ gaps, together with that from the full-event method given in Fig.~\ref{fig:fcmeampt}. The $\Delta\eta$ values are the $\eta$ gap between the two sub-events that are symmetric about midrapidity. 
We have used $\lambda_\ampt=15\%$ and scaled the obtained $\enf$ from \ampt\ by the same factor used to match the $\enf^{\rm exp}$ in the full-event method.
Once an $\eta$ gap is applied, \ampt\ gives significantly reduced $\fcme^*$ values because of the significantly reduced nonflow $\enf$ contamination (note that the average inter-particle $\eta$ gap is significantly larger than the $\Delta\eta$ value between the sub-events).
The $\fcme^*$ resulting from \hijing\ 3p nonflow is, however, not much reduced. This is consistent with the fact that the 3p nonflow in \hijing\ is primarily due to di-jet correlations which are not much affected by the $\eta$ gap. As a result, the sub-event method gives an overall negative $\fcme^*$, approximately $\fcme^*=(-5\pm3)\%$.

\begin{figure}
	\includegraphics[width=1.0\linewidth]{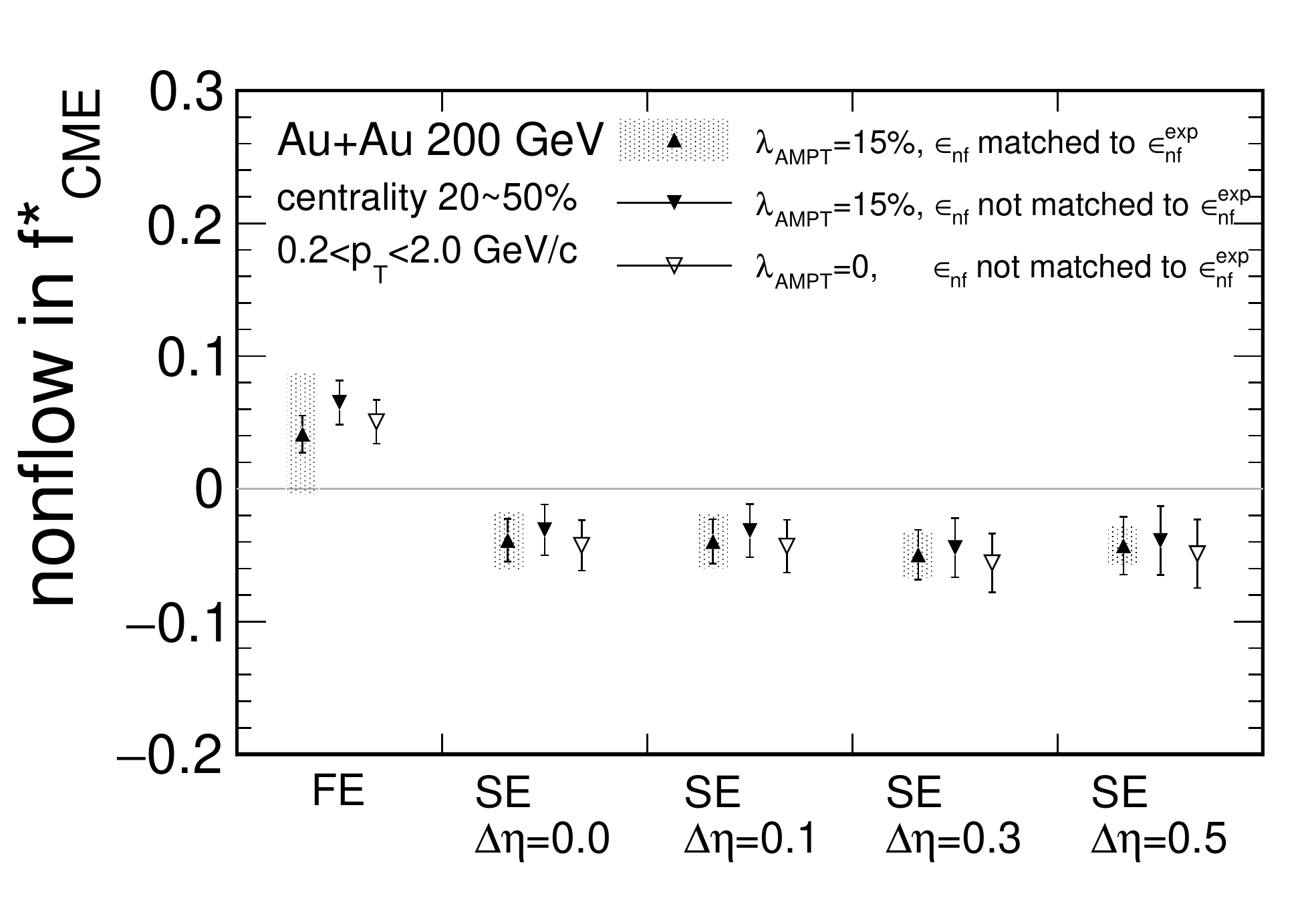}
	\caption{The nonflow $\fcme^*$ in 20-50\% Au+Au collisions obtained with various estimates of $\enf$: from \ampt\ via Eq.~(\ref{eq:lambda}) with $\lambda_{\ampt}=15\%$ and scaled to data $\enf^{\rm exp}$~\cite{Abdelwahab:2014sge} (i.e.~solid triangles from Fig.~\ref{fig:fcme}), and without scaled to data, one with $\lambda_{\ampt}=15\%$ (inversed solid triangles) and the other $\lambda_{\ampt}=0\%$ (inversed open triangles). The \poi\ and particle $c$ are from $|\eta|<1$ and  $0.2<p_T<2$~GeV/$c$.}
	\label{fig:syst}
\end{figure}

The largest uncertainty of our nonflow estimates comes from those on the experimental nonflow $\enf^{\rm exp}$ measurements~\cite{Abdelwahab:2014sge}. To give another assessment, we show in Fig.~\ref{fig:syst} the nonflow estimates taking $\enf=\epsilon_{\ampt}$ directly from \ampt\ (as shown in Fig.~\ref{fig:ampt}(d)), without the multiplicative factor to match to data $\enf^{\rm exp}$. Both $\lambda_{\ampt}=15\%$ and 0\% results are shown (their difference is insignificant)
where the error bars are statistical as from the models. These are generally within the systematic uncertainties of our estimates in the solid triangles, indicating the robustness of our estimates.

\begin{figure}
	\includegraphics[width=1.0\linewidth]{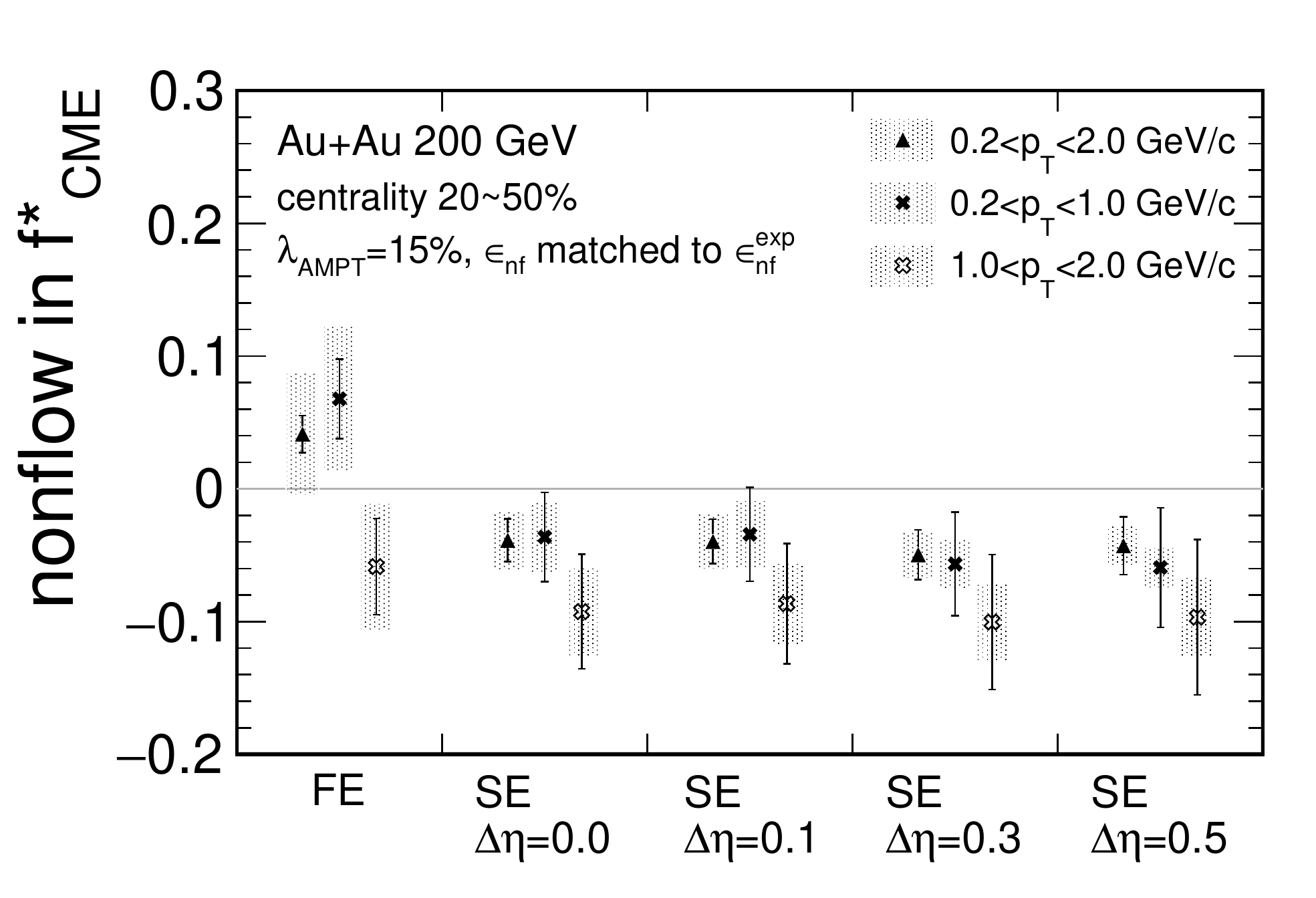}
	\caption{Same as the solid triangles in Fig.~\ref{fig:fcme}, but with two additional sets of data points of split $p_T$ ranges for the \poi\ and particle $c$: $0.2<p_T<1$~GeV/$c$ (filled crosses) and $1<p_T<2$~GeV/$c$ (open crosses).}
	\label{fig:pt}
\end{figure}
\begin{figure}
	\includegraphics[width=1.0\linewidth]{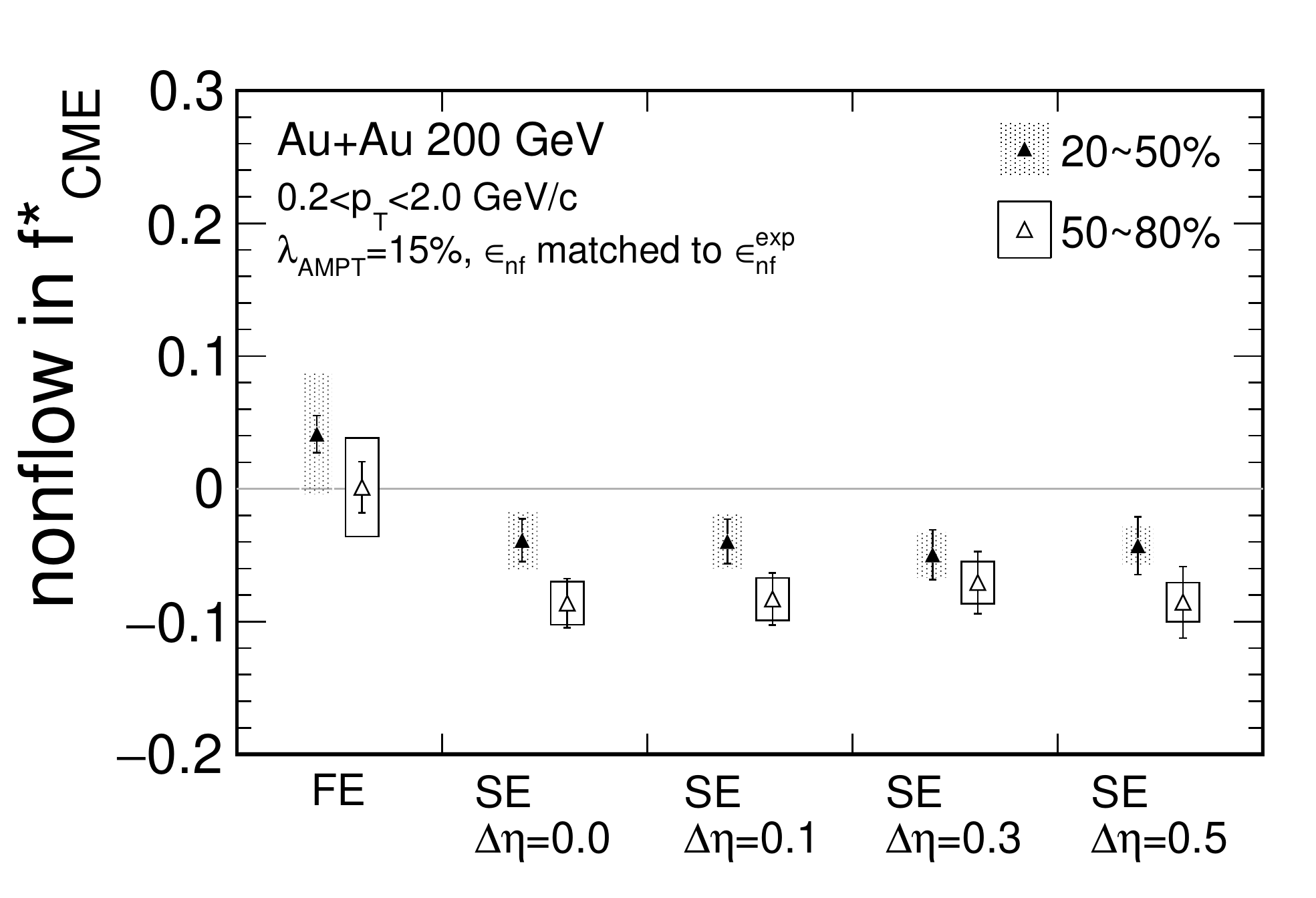}
	\caption{The nonflow $\fcme^*$ estimates in 200~GeV Au+Au collisions for 20-50\% centrality (solid triangles, as same as those in Fig.~\ref{fig:fcme}) and 50-80\% centrality (open triangles).}
	\label{fig:peri}
\end{figure}

Nonflow has strong $p_T$ dependence; di-jet correlations are more significant at high $p_T$ as modeled in \hijing. 
We repeat our analysis separating the \poi\ (and particle $c$) into two $p_T$ bins: $0.2<p_T<1$~GeV/$c$ and $1<p_T<2$~GeV/$c$ (with the same $|\eta|<1$ range). The results are shown in Fig.~\ref{fig:pt} by the crosses. The 2p nonflow is taken from \ampt\ scaled by the centrality-dependent parameterization from the $0.2<p_T<2.0$~GeV/$c$ range in Fig.~\ref{fig:FitScl}. It may be reasonable at low $p_T$, but unlikely correct at high $p_T$ as \ampt\ destroys minijet correlations at its initialization. The 3p nonflow which we take from \hijing\ should be reasonable at high $p_T$ and may likely be so at low $p_T$ as well. Nevertheless, the nonflow effect for $0.2<p_T<1$~GeV/$c$ is similar to that for $0.2<p_T<2$~GeV/$c$.
We observe a more negative $\fcme^*$ for $1<p_T<2$~GeV/$c$ although the  statistical uncertainties are significantly larger (note the full $p_T$ range contains more statistics than the sum of the two individual $p_T$ ranges because of cross $p_T$ range pairs). The larger negative $\fcme^*$ is mainly caused by the increased negative effect at high $p_T$ from di-jet 3p correlations.

We repeat our analysis of Fig.~\ref{fig:fcme} for peripheral 50-80\% collisions. The results are shown in Fig.~\ref{fig:peri} by the open triangles together with those of the 20-50\% centrality range from Fig.~\ref{fig:fcme}. The results in peripheral collisions are systematically shifted towards more negative $\fcme^*$ compared to central collisions. This is mainly due to a more significant 3p correlation effect.

\begin{figure}
	\includegraphics[width=1.0\linewidth]{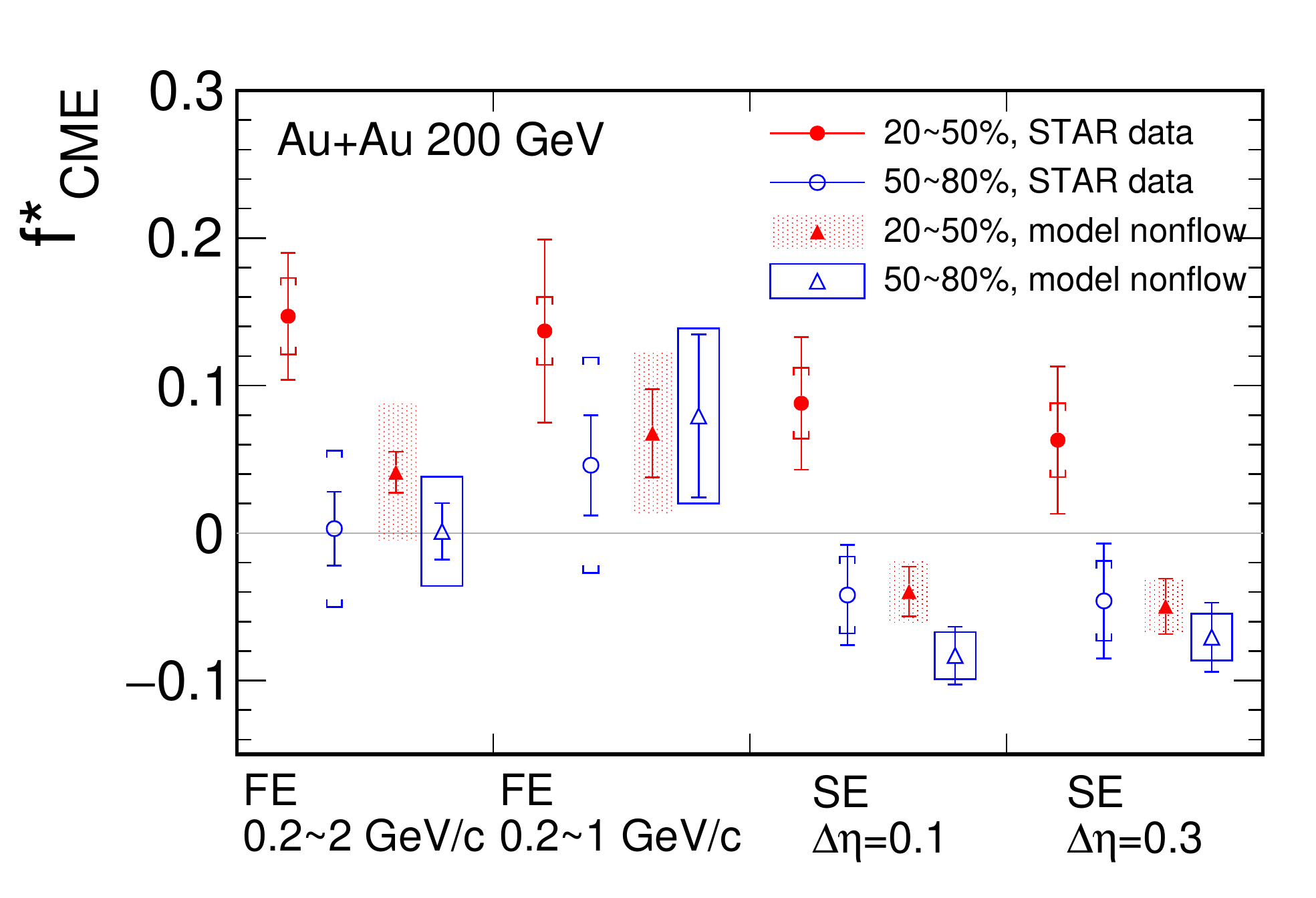}
	\caption{STAR measurements~\cite{Abdallah:2021itw} of $\fcme^*$ together with our nonflow estimates within 20--50\% (filled markers) and 50-80\% (open markers) centralities in 200~GeV Au+Au collisions. The \poi\ and particle $c$ (for \sep) are from $|\eta|<1$ and $0.2<p_T<2$~GeV/$c$ except the second set of points where $0.2<p_T<1$~GeV/$c$. 
	The 2p nonflow effect $\enf$ is matched to data $\enf^{\rm exp}$~\cite{Abdelwahab:2014sge} with $\lambda_{\ampt}=15\%$, and the 3p nonflow contribution is $\epsilon_3=\epsilon_3^{\hijing}\pm20\%$~\cite{Zhao:2019kyk}.}
	\label{fig:data}
\end{figure}

STAR has measured the $\fcme^*$ using the \ssp/\spp\ method in Au+Au collisions at 200~GeV~\cite{Abdallah:2021itw}. 
We tabulate the STAR measurements together with our estimates of nonflow contributions in Table~\ref{tab}.
We compare them in Fig.~\ref{fig:data} where the STAR data are shown by the red stars and our nonflow estimates are shown by the black triangles. 
The peripheral collision data are mostly consistent with our nonflow estimates. The central collisions data are systematically larger than our estimations of nonflow contributions (except for the low $p_T$ results).
If our nonflow estimations are robust, then the STAR measurements seem to suggest finite \cme\ signals. 

\begin{table*}[hbt]
    \caption{The STAR \ssp/\spp\ measurements of $\fcme^*$ by the full-event (FE) and sub-event (SE) methods~\cite{Abdallah:2021itw} together with our estimated nonflow contributions in 20-50\% Au+Au collisions at $\snn=200$~GeV. The estimates assume charge-dependent 2p correlation effect $\epsilon_2=1.7\epsilon_2^{\ampt}$~\cite{Zhao:2019kyk}, charge-dependent 3p correlation effect $\epsilon_3=\epsilon_3^{\hijing}\pm20\%$~\cite{Zhao:2019kyk}, and charge-independent 2p nonflow $\enf$ by Eq.~(\ref{eq:lambda}). Two cases of $\enf$ are tabulated: matched to the experimental data~\cite{Abdelwahab:2014sge} 
    with $\lambda_{\ampt}=15\%$ (the $\lambda_{\ampt}=0\%$ results are similar), and not matched to data with $\lambda_{\ampt}=0\%$ (the $\lambda_{\ampt}=15\%$ results are similar). The first (or only) quoted error is statistical and the second systematic.}
    \centering
    \begin{tabular}{c|cccc}\hline\hline
    & FE ($p_T$=0.2-2~GeV/$c$) & FE ($p_T$=0.2-1~GeV/$c$) & SE ($\Delta\eta=0.1$) & SE ($\Delta\eta=0.3$) \\ \hline
    STAR data & $(14.7\pm 4.3\pm 2.6)\%$ & $(13.7\pm 6.2\pm 2.3)\%$ & $(8.8\pm 4.5\pm 2.4)\%$ & $(6.3\pm 5.0\pm 2.5)\%$ \\ \hline
    $\enf$ matched to $\enf^{\rm exp}$, $\lambda_\ampt=15\%$ & $\fullnumber$ & $\fullnumberlowpt$ & $\subnumber$ & $\sssnumber$ \\
    %$\enf^{\rm exp}=(10\pm4)\%$ & $\tenfullnumber$ & $\tensubnumber$ & $\tensssnumber$ \\
    %$\enf$ not matched to $\enf^{\rm exp}$, $\lambda_\ampt=15\%$ & $\nosclfullnumber$ & $\nosclsubnumber$ & $\nosclsssnumber$ \\
    $\enf$ not matched to $\enf^{\rm exp}$, $\lambda_\ampt=0\%$ & $\simufullnumber$ & $\simufullnumberlowpt$ &  $\simusubnumber$ & $\simusssnumber$ \\
    %$\enf$ not matched to $\enf^{\rm exp}$, $\lambda_\ampt=0\%$, $1.0<p_{T}<2.0 \text{ GeV/c}$ & $\simufullnumberhighpt$ & $\simusubnumberhighpt$ & $\simusssnumberhighpt$ \\
    \hline\hline
    \end{tabular}
    \label{tab}
\end{table*}

%------------------------------------------------------------------------------------------------%

\section{Summary and Outlook}\label{sec:summary}

The 3p azimuthal correlator $\dg^*$ is dominated by the flow-induced charge-dependent 2p correlation background. The \ssp/\spp\ method~\cite{Xu:2017qfs} has been proposed to extract the \cme\ signal fraction, $\fcme^*$, in the measured $\dg^*$ by assuming the background to be proportional to the measured elliptic flow $v_2^*$. 
The charge-independent 2p nonflow contamination in $v_2^*$ and the charge-dependent 3p nonflow contribution to $\dg^*$ are two further background sources in the extracted $\fcme^*$.
In this paper we have investigated the effects of these nonflow backgrounds. It is shown that the effects from 2p and 3p nonflow correlations in $\fcme^*$ are opposite in sign. They partially cancel each other, making the $\fcme^*$ less vulnerable to nonflow.
The \ampt\ and \hijing\ models are used, together with constraints from experimental data, to quantitatively estimate the magnitudes of those nonflow effects. The main result is given by Eq.~(\ref{eq:fcmedata}) and Fig.~\ref{fig:fcme}. 
The main ingredients of our estimation are as follow.
\begin{itemize}
    \item \ampt\ contains mainly 2p correlations and thus the calculated $\fcme^*$ gives a good estimate of the $v_2^*$ nonflow in \ampt, $\enf=\epsilon_{\ampt}$. The $\enf$ from \ampt\ is scaled to match the experimentally deduced nonflow $\enf^{\rm exp}$~\cite{Abdelwahab:2014sge}. 
    The residual charge-dependent 3p correlations in \ampt, on the order $\lambda_\ampt\approx15\%$ of those in \hijing~\cite{Zhao:2019kyk}, have negligible effect in our estimation. % of the 2p nonflow contribution to $\fcme^*$.
    \item Charge-dependent 3p correlations $\epsilon_3$ are the main nonflow contribution to $\dg^*$. The charge-dependent 3p (di-jet) correlations in \hijing, $\epsilon_3^{\hijing}$, are found to give a fair description of the experimental data in small-system collisions ($\epsilon_3=\epsilon_3^{\hijing}\pm20\%$)~\cite{Zhao:2019kyk}. They are used, together with $\epsilon_2=1.7\epsilon_2^{\ampt}$, the flow-induced background $\dg_{\rm bkgd}$ in \ampt\ scaled to match data measurement of $\dg^*$, to estimate the 3p nonflow contribution to $\fcme^*$.
\end{itemize}

It is found, with 2p and 3p nonflow correlations in \ampt\ and \hijing\ together with constraints from experimental data, 
that the nonflow contribution 
in 20--50\% centrality Au+Au collisions at $\snn= 200$~GeV with the full-event method (without pseudorapidity gap) is 
approximately $(4\pm5)\%$.
With the sub-event method with pseudorapidity gaps,
the nonflow contribution to $\fcme^*$ is generally negative, approximately $(-5\pm3)\%$.  
The implications of our nonflow estimates to the STAR measurements are highlighted in Table~\ref{tab} and Fig.~\ref{fig:data}.
They suggest that the STAR measurements 
may imply a finite \cme\ signal.

Further reduction in 2p nonflow contamination in $v_2^*$ by applying larger $\eta$ gaps, for example, via the forward event-plane detector~\cite{Adams:2019fpo} newly installed in STAR, would be desirable. A forward \sep\ or particle $c$ will, in addition, enable reasonable $\eta$ gaps to be applied also between the midrapidity $\alpha$ and $\beta$ particles, reducing the flow-induced background $\dg_{\rm bkgd}$ and 3p nonflow contaminations. Future Au+Au runs by STAR with the enhanced forward capability and expected large data volumes would provide definite conclusion on the \cme.

Isobar collision data have been collected by STAR in 2018~\cite{Skokov:2016yrj} and blind data analysis is ongoing~\cite{Adam:2019fbq}. An unambiguous (relative) \cme\ signal may emerge from these data, dependent of the signal strength given by Mother Nature~\cite{Deng:2016knn}. A recent estimate using the Anomalous-Viscous Fluid Dynamics prediction of the \cme\ strength in isobar collisions suggests an effect only on the order of $2\sigma$ significance~\cite{Feng:2021oub}. 
In any case, the absolute magnitude of the possible \cme\ signal would have large uncertainty, which would require large-volume Au+Au collision data to resolve. Thus, regardless of the isobar data outcome, future heavy ion runs by STAR will be important for the \cme\ physics.

%------------------------------------------------------------------------------------------------%

\section*{Acknowledgments}

This work is supported by the U.S.~Department of Energy (Grant No.~DE-SC0012910), 
the China Hubei Province Department of Education (Grant No.~D20201108), 
the China National Natural Science Foundation (Grant Nos.~11905059, 12035006, 12075085, 12047568), and 
the China Ministry of Science and Technology (Grant No.~2020YFE0202001).

%------------------------------------------------------------------------------------------------%

\bibliography{./ref}

%------------------------------------------------------------------------------------------------%

\end{document}